\documentclass[twocolumn]{aastex701}
	
\usepackage{color}
\usepackage{amsmath}
\usepackage{multirow}

\shorttitle{Paschen-based Black Hole Mass Estimators and Luminosity Functions}
\shortauthors{Jiang et al.}

\begin{document}
\title{COSMOS-3D: Black Hole Mass Estimators and Luminosity Functions of Paschen-line AGNs}

\author[0009-0003-6747-2221]{Danyang Jiang}
\email{jiangdy@stu.pku.edu.cn}
\affiliation{Department of Astronomy, School of Physics, Peking University, Beijing 100871, China}
\affiliation{Kavli Institute for Astronomy and Astrophysics, Peking University, Beijing 100871, China}
\author[0000-0003-4176-6486]{Linhua Jiang}
\email{jiangkiaa@pku.edu.cn}
\affiliation{Department of Astronomy, School of Physics, Peking University, Beijing 100871, China}
\affiliation{Kavli Institute for Astronomy and Astrophysics, Peking University, Beijing 100871, China}

\author[0000-0003-0964-7188]{Shuqi Fu}
\email{fushuqi@stu.pku.edu.cn}
\affiliation{Department of Astronomy, School of Physics, Peking University, Beijing 100871, China}
\affiliation{Kavli Institute for Astronomy and Astrophysics, Peking University, Beijing 100871, China}

\author[0000-0002-2420-5022]{Zijian Zhang}
\email{zjz.kiaa@stu.pku.edu.cn}
\affiliation{Department of Astronomy, School of Physics, Peking University, Beijing 100871, China}
\affiliation{Kavli Institute for Astronomy and Astrophysics, Peking University, Beijing 100871, China}

\author[0000-0002-4765-1500]{Jie Chen}
\email{jiechen@stu.pku.edu.cn}
\affiliation{Department of Astronomy, School of Physics, Peking University, Beijing 100871, China}
\affiliation{Kavli Institute for Astronomy and Astrophysics, Peking University, Beijing 100871, China}

\author[0000-0003-0230-6436]{Zhiwei Pan}
\email{zhiweip@illinois.edu}
\affiliation{Department of Astronomy, University of Illinois Urbana-Champaign, Urbana, IL 61801, USA}

\author[0000-0002-1234-552X]{Shengxiu Sun}
\email{sxsun@stu.pku.edu.cn}
\affiliation{Department of Astronomy, School of Physics, Peking University, Beijing 100871, China}
\affiliation{Kavli Institute for Astronomy and Astrophysics, Peking University, Beijing 100871, China}
\affiliation{Institute for Theoretical Physics, Heidelberg University, Philosophenweg 12, 69120, Heidelberg, Germany}

\author[0000-0002-4622-6617]{Fengwu Sun}
\email{fengwu.sun@cfa.harvard.edu}
\affiliation{Center for Astrophysics $|$ Harvard \& Smithsonian, 60 Garden St., Cambridge, MA 02138, USA}

\author[0000-0001-6947-5846]{Luis C. Ho}
\email{lho.pku@gmail.com}
\affiliation{Kavli Institute for Astronomy and Astrophysics, Peking University, Beijing 100871, China}
\affiliation{Department of Astronomy, School of Physics, Peking University, Beijing 100871, China}

\author[0000-0002-4569-9009]{Jinyi Shangguan}
\email{shangguan@pku.edu.cn}
\affiliation{Kavli Institute for Astronomy and Astrophysics, Peking University, Beijing 100871, China}
\affiliation{Department of Astronomy, School of Physics, Peking University, Beijing 100871, China}

\author[0000-0002-9382-9832]{Andreas L. Faisst}
\email{afaisst@ipac.caltech.edu}
\affiliation{IPAC, California Institute of Technology, 1200 E. California Blvd. Pasadena, CA 91125, USA}

\author[0009-0004-2538-7237]{Olivier Gilbert}
\email{ogilbert@umich.edu}
\affiliation{Department of Astronomy, University of Michigan, 1085 S. University Ave., Ann Arbor, MI 48109, USA}

\author[0000-0001-6251-649X]{Mingyu Li}
\email{lmy22@mails.tsinghua.edu.cn}
\affiliation{Department of Astronomy, Tsinghua University, Beijing 100084, China}

\author[0000-0003-4247-0169]{Yichen Liu}
\email{yichenliu@arizona.edu}
\affiliation{Steward Observatory, University of Arizona, 933 N Cherry Avenue, Tucson, AZ 85721, USA}

\author[0000-0001-7634-1547]{Zi-Jian Li}
\email{zjli@nao.cas.cn}
\affiliation{Chinese Academy of Sciences South America Center for Astronomy, National Astronomical Observatories of China,\\ CAS, 20A Datun Road, Beijing 100012, China}
\affiliation{School of Astronomy and Space Sciences, University of Chinese Academy of Sciences, Beijing 100049, China}

\author[0009-0003-4742-7060]{Takumi S. Tanaka}
\email{takumi.tanaka@ipmu.jp}
\affiliation{Department of Astronomy, Graduate School of Science, The University of Tokyo, 7-3-1 Hongo, Bunkyo-ku, Tokyo, 113-0033, Japan}
\affiliation{Kavli Institute for the Physics and Mathematics of the Universe (WPI), The University of Tokyo Institutes for Advanced Study, The University of Tokyo, Kashiwa, Chiba 277-8583, Japan}
\affiliation{Center for Data-Driven Discovery, Kavli IPMU (WPI), UTIAS, The University of Tokyo, Kashiwa, Chiba 277-8583, Japan}

\begin{abstract}

Near-IR Paschen lines are potentially an excellent tracer of Type 1 AGNs that is hardly affected by dust extinction. JWST allows us, for the first time, to explore Paschen-line objects at redshift $z>1$. Here we present a study of 62 AGNs with broad Pa$\alpha$ and Pa$\beta$ lines at $1<z<3$ using data from the JWST COSMOS-3D program. These AGNs are efficiently selected and identified using NIRCam imaging and grism slitless spectroscopic data. 
We separate the AGN-host emission with image decomposition and quantify dust attenuation with multi-band data.
We construct a calibration sample with optical spectroscopy and use single-epoch, \ion{Mg}{2}-based black hole masses ($M_{\mathrm{BH}}$) as an anchor to derive new, Paschen-based $M_{\mathrm{BH}}$ estimators.  
We obtain three sets of $M_{\mathrm{BH}}$ estimators based on Paschen line luminosities and AGN continuum luminosities at 1 and 2~$\mu$m, respectively. 
After dust corrections, they are well consistent with each other, and also broadly agree with previous results. 
With this AGN sample, we further construct the first Pa$\alpha$ and Pa$\beta$ luminosity functions (LFs) of Type 1 AGNs. The derived LFs are $3-5$ times higher than those of UV/optical-selected AGNs, indicating that Paschen-selected Type 1 AGNs are more complete. In addition, the intrinsic properties of our AGNs show no dependence on dust reddening, suggesting that the observed reddening is unrelated to the central engine and is thus likely caused by line-of-sight obscuration. 

\end{abstract}


\graphicspath{{./}}

\section{Introduction} \label{sec:introduction}

Supermassive black holes (SMBHs) are broadly accepted to power active galactic nuclei (AGNs) by releasing gravitational potential energy during matter accretion. The discovery of tight correlations between SMBH masses and host galaxy properties indicates the important role of SMBHs in galaxy evolution \citep[e.g.,][]{gebhardt2000,ferrarese2000,marconi2003,gultekin2009,kormendy2013,abuter_2024}. SMBH mass measurement is therefore fundamental for understanding the co-evolution of SMBHs and their host galaxies. Gas in the AGN broad-line region (BLR) is assumed and also observationally supported \citep[e.g.,][]{Peterson2004} to virially orbit around SMBHs, producing Doppler-broadened emission lines. In this scenario, the virial theorem can be applied to estimate the black hole mass: 
\begin{equation} 
M_{\mathrm{BH}}=f\frac{R_{\mathrm{BLR}}\Delta V^2}{G},
\end{equation}
where $R_{\mathrm{BLR}}$ is the BLR size, $f$ is a factor related to the geometry and kinematics of the BLR, and $G$ is the gravitational constant. The velocity dispersion $\Delta V$ is usually represented by the full width at half-maximum (FWHM) of the broad emission line. 

Reverberation mapping (RM) measures the time delay between the continuum and emission-line variation, and is a standard method for measuring $R_{\mathrm{BLR}}$ (see \citealt{peterson1993a,Peterson2004,cackett2021} for reviews). In the virial framework, the combination of $R_{\mathrm{BLR}}$ and the broad-line width provides an estimate of the SMBH mass. Recently, near-IR interferometry with VLTI/GRAVITY has enabled spatially resolved measurements of the BLR structure and kinematics, first in local and low-redshift AGNs and quasars \citep{GRAVITY2018,GRAVITY2020,GRAVITY2021}, and subsequently in increasingly distant systems with the improved sensitivity of GRAVITY+ \citep{abuter_2024,Gravity2026}.

Both RM and interferometric measurements rely on substantial observational resources. 
Based on the RM measurements, single-epoch methods have been developed as a more efficient approach to estimate $M_{\mathrm{BH}}$ \citep[e.g.,][]{vestergaard2002c}. 
In these methods, $R_{\mathrm{BLR}}$ is inferred from the continuum or line luminosity based on the empirical relation between $R_{\mathrm{BLR}}$ and luminosity (hereafter the R–L relation) \citep[e.g.,][]{shuder1981,kaspi2005a,bentz2013,du2019}. 
The interferometric measurements also broadly consistent with the classical RM-based R-L relation \citep[e.g.,][]{GRAVITY2024}.
After calibration, the single-epoch $M_{\mathrm{BH}}$ estimators can be easily applied to large AGN/quasar samples.

Single-epoch $M_{\mathrm{BH}}$ estimators have been established across a wide wavelength range. In the rest-frame ultraviolet (UV), the popular $M_{\mathrm{BH}}$ estimator is based on the monochromatic luminosity $L_{3000}$ ($\lambda L_\lambda$) at rest-frame 3000 \AA\ and the \ion{Mg}{2} line width \citep[e.g.,][]{vestergaard2009,wang2009,woo2018a,le2020,shen2024,pan2025b,bai2025}. Another UV indicator uses the \ion{C}{4} line width with the monochromatic luminosity $L_{1350}$. It has been known that the \ion{C}{4}-based $M_{\mathrm{BH}}$ estimator exhibits a substantially larger scatter, even after correction for biases introduced by outflows (e.g., \citealt{coatman2017}). In the rest-frame optical, the Balmer emission lines H$\alpha$ and H$\beta$, together with the continuum luminosity at 5100\AA\ ($L_{5100}$) or the line luminosity, provide widely used $M_{\mathrm{BH}}$ estimators (e.g., \citealt{kaspi2000,greene2005,grier2017,shen2024,wang2024b}). Extending to the rest-frame near-infrared (NIR), hydrogen Paschen lines, such as Pa$\alpha$ $\lambda$1.8756$\mu$m and Pa$\beta$ $\lambda$1.2821$\mu$m, arise from transitions from higher energy levels to the $n=3$ state. The GRAVITY measurement of the Pa$\alpha$ BLR provides $M_{\mathrm{BH}}$ that are well consistent with the RM results of H$\beta$ \citep{sturm2018a,Zhang2019}.

NIR Paschen lines offer distinct advantages comparing to the rest-frame UV and optical lines. NIR lines are less affected by dust attenuation, so Paschen-based $M_{\mathrm{BH}}$ estimators can be applied to both blue and dust-reddened AGNs, enabling reliable $M_{\mathrm{BH}}$ measurements even when the optical and UV broad lines are unavailable. In addition, Type 1 AGNs selected via broad Paschen lines are less biased by dust attenuation and thus more complete than UV and optical selected samples. Therefore, broad Paschen-line AGNs provide a more comprehensive view of the Type 1 AGN population.

The exploration of Paschen-line AGNs has been limited by the challenge of NIR observations. Under the Case B assumption, Pa$\alpha$ is about eight times fainter than the corresponding H$\alpha$, making it difficult to detect. In addition, Paschen lines fall in wavelength ranges that are strongly affected by atmospheric absorption. Consequently, ground-based NIR spectroscopic studies have been confined to small samples of Paschen-line AGNs in the local Universe \citep{landt2008,glikman2007a,glikman2012,kim2015}. A few studies have calibrated Paschen-based $M_{\mathrm{BH}}$ estimators using these local samples \citep{kim2010,landt2011,landt2013}. Furthermore, the host-galaxy contribution becomes non-negligible in the rest-frame NIR, necessitating high–resolution imaging to reliably quantify and remove host galaxy in studies of AGN properties.

James Webb Space Telescope (JWST) provides the first opportunity to investigate Paschen-line AGNs beyond the local Universe. The powerful spectroscopy of JWST NIRCam/Grism and NIRSpec have yielded a large number of Paschen lines at cosmic noon \citep[e.g.,][]{oesch2023,zhuang2024,curtis-lake2025,sun2025}. Meanwhile, the high-resolution, multiple-band NIRCam images enables a reliable decomposition of AGN host galaxies \citep[e.g.,][]{ding2022,zhuang2024b,tanaka2025a,chen2025f,jiang2025c,li2025b}. With the rapidly growing JWST Paschen-line database, Paschen-based $M_{\mathrm{BH}}$ estimators are becoming increasingly important. Furthermore, the high completeness of JWST NIRCam/Grism samples will allow a robust determination of Paschen-based AGN luminosity functions (LFs).

In this paper, we will use JWST to build a complete Type 1 AGN sample with broad Pa$\alpha$ and Pa$\beta$ at cosmic noon, and develop new, single-epoch $M_{\mathrm{BH}}$ estimators based on the Paschen lines. We will also calculate the LFs of the AGN sample, and analyze their black hole properties. The structure of the paper is as follows. We introduce our datasets and the sample selection in Section \ref{sec:data}. We describe our spectral fitting, host galaxy decomposition and spectral energy distribution (SED) fitting in Section \ref{sec:analyses}. The resultant $M_{\mathrm{BH}}$ estimators and Paschen-line luminosity functions are shown in Section \ref{sec:results}. We compare our results with previous works and provide some implications for dusty Type 1 AGNs in Section \ref{sec:discussion}. The paper is summarized in Section \ref{sec:summary}. We adopt a standard $\Lambda$CDM cosmology with $H_0$=70 km s$^{-1}$ Mpc$^{-1}$, $\Omega_\mathrm{m}$=0.3, and $\Omega_\mathrm{\Lambda}$=0.7. All the magnitudes are reported in the AB system \citep{Oke1983}.

\section{Data and Sample Construction} \label{sec:data}

\subsection{JWST/NIRCam Grism Data and Reduction} \label{jwst_spec_data}

Our data are primarily from JWST COSMOS-3D program. COSMOS-3D is a 268 hour JWST Cycle 3 Large Program (GO-5893) that observes 0.33 deg$^2$ in the COSMOS field with NIRCam Wide Field Slitless Spectroscopy (WFSS) and broad-band NIRCam and MIRI imaging (Kakiichi et al. in preparation). NIRCam WFSS observations use the row-direction grism (Grism R) combined with the wide-band filter F444W. With its large areal coverage, COSMOS-3D yields a large sample of emission-line sources from the local Universe to high redshift that cover strong lines such as Br$\alpha$, Br$\beta$, Pa$\alpha$, Pa$\beta$, \ion{He}{1}, [S III], H$\alpha$, H$\beta$, and [O III] \citep[Fu et al. in preparation, Wang et al. in preparation,]{meyer2025a,lin2025b}. Pa$\alpha$ and Pa$\beta$ at $z\sim1-3$ account for a significant fraction of the emission-line sources. The spectral resolution of WFSS ($R\sim$1600) allow us to robustly identify broad-line AGNs. We constructed the broad Pa$\alpha$ and Pa$\beta$ AGN sample based on the COSMOS-3D WFSS data.

The WFSS data were reduced following the standard procedure \citep{sun_2023,Fu2025}. The corresponding codes and calibration files are publicly available\footnote{\url{https://github.com/fengwusun/nircam_grism}}. We first processed the NIRCam WFSS data using the standard JWST Stage-1 calibration pipeline (version \verb|1.13.4|) and the calibration reference file \verb|jwst_1364.pmap|. A flat-field correction was applied using flat-field data obtained with the same filter and detector. For each exposure, we constructed and subtracted a super-sky background using all flat-fielded exposures taken with the same filter+pupil+module combination. We then performed an additional 2D sky-background subtraction using the Source Extractor algorithm to further remove the residual \citep{Bertin_1996}. The background was estimated on the 2.5$\sigma$-clipped image using a mode estimator of the form (2.5 $\times$ median) - (1.5 $\times$ mean), which is less sensitive to source crowding than a simple clipped mean. If (mean - median)/STD $>$ 0.3, where STD denotes the standard deviation, the mode estimator becomes unreliable and a simple median is used instead. Finally, we refined the astrometry of the WFSS data by matching the simultaneously obtained short-wavelength images to the COSMOS2025 source catalog \citep{COSMOS2025_cat}, which was registered to the HST reference frame.

\subsection{JWST/NIRCam Image and Photometry} \label{jwst_img_data}

We used a complete set of JWST/NIRCam imaging available in the COSMOS field, including F115W, F150W, F200W, F277W, F356W, and F444W bands. The images are mainly from proposal IDs 1635, 1727, 1810, 1837, 1933, 2321, 2514, 3990, 5398, 5893, 6368, 6434, and 6585. We reduced the data using the combination of the JWST Calibration Pipeline (v.1.17.0), the scripts for NIRCam imaging reduction for the Cosmic Evolution Early Release Science (CEERS) Survey \citep{Bagley2023}, and our own custom codes. The reduced JWST images were used to decompose the host galaxy in Section \ref{decompse}.

We adopted the JWST photometric data provided by Fu et al. (in preparation). We also used images from Hubble Space Telescope (HST) F814W dataset \citep{koekemoer2007}, and Ultradeep optical imaging data in the $g,r,i,z,y$ broad bands from the Hyper Suprime-Cam (HSC) Subaru Strategic Program \citep{Aihara2022}. For the JWST and HST images, we constructed empirical point spread function (PSF) models for each band using \texttt{PSFEx} \citep{bertin2013}. We then performed photometry using \texttt{SExtractor} \citep{Bertin_1996} on the PSF-matched images with Kron-aperture correction. For the HSC images, we derived template-fitting deblending photometry to get the total fluxes at each band with T-PHOT v2.0 \citep{Merlin2016}. The detailed procedure is presented in Fu et al. (in preparation, see also \citealt{fu2025d}).

\subsection{Selection of Broad Paschen-line Sample} \label{selection}

We systematically searched for broad line AGNs in the COSMOS-3D field.  
We started with compact bright sources with F444W Kron magnitude brighter than 23 mag and compactness (flux $_{r=0\farcs3}$ / flux$_{r=0\farcs15}$) $<$ 1.8 in the COSMOS2025 catalog. 
This compactness selection removes extended sources that produce morphologically broadened lines in the grism spectra.
While the compactness threshold is not sufficient to remove all extended sources, it ensures that known point-source AGNs are not excluded. We also included the saturated point sources that were rejected by the compactness criterion. We extracted the 2D spectra and calibrated the fluxes using the grism sensitivity models. To detect broad emission lines, we further extracted 1D spectra using a fixed box in five pixels. We first subtracted a median-filtered continuum model from the 1D spectra. We then detected emission lines with integrated signal-to-noise ratios (SNRs) greater than 3 on the resulting line spectra to obtain line centers. Because the extended wings of broad lines can be over-subtracted by the median-filtered continuum, we subsequently fitted a linear continuum around each detected line center in the 1D spectra and re-subtracted the continuum. Finally, each line was fitted with two models:
\begin{itemize}
    \item[(1)] A single-Gaussian model with FWHM $<$ 12000 $\rm km~s^{-1}$;
    \item[(2)] A double-Gaussian model with one narrow profile with FWHM $<$ 500 $\rm km~s^{-1}$ plus one broad profile with 500 $\rm km~s^{-1}$ $<$ FWHM $<$ 12000 $\rm km~s^{-1}$.
\end{itemize}
We adopted 12000 $\rm km~s^{-1}$ as a reasonable upper limit for real AGN broad lines. In the double-Gaussian model, we restricted the narrow component to FWHM $<$ 500 $\rm km~s^{-1}$ to prevent it from contaminating the identification of the broad component. These criteria were used only for the sample selection. 
All the FWHMs measured in this work were limited to be greater than the resolved FWHM. The line fitting was performed using the least-square method. The model with the smaller Bayesian Information Criterion (BIC) parameter was determined as the best-fit model. The BIC parameter is defined as $\mathrm{BIC} = \chi^2 + k \mathrm{ln}n$ , where $k$ is the number of free parameters and $n$ is the number of data points \citep{liddle2007a}. 
Emission lines with one FWHM $>$ 1000 $\rm km~s^{-1}$ in the best-fit model were selected as initial broad lines.

After the above selection, we visually inspected all selected sources to remove the contamination, identified real lines, and determined their redshifts. For each source, we listed all possible contamination sources aligning in the same dispersion region of the grism spectrum. We simultaneously considered all the information, including photometric redshifts, SED shapes, multiple-line locations, line morphologies, and existing spectral redshifts. We associated each detected broad emission line with its true emitting source, identified emission lines, and determined redshifts. The detailed description of this visual-inspection procedure will be presented in Fu et al. (in preparation).  

We further refined the identified Pa$\alpha$ and Pa$\beta$ sample. We visually inspected their F444W images and excluded sources with obvious extended features that contaminated the line profiles and were not removed by the compactness criterion. We reprocessed the remaining sample with improved spectral extraction and more accurate broad-line measurements to remove contaminants. We optimally extracted 1D spectra following \cite{horne1986} and subtracted 1D continuum by fitting a linear model. The fitting windows were individually adjusted to minimize the contamination flux from nearby sources. The continuum of ID50403 is largely blended with the continuum of a nearby source. We performed a simultaneous two-component modeling of the continua in the 2D spectra using Gaussian tracing functions and subtracted the contaminating component. The 1D spectrum of ID50403 was then extracted from the corrected 2D spectrum.

\begin{figure*}[t]
\epsscale{1.2}
\plotone{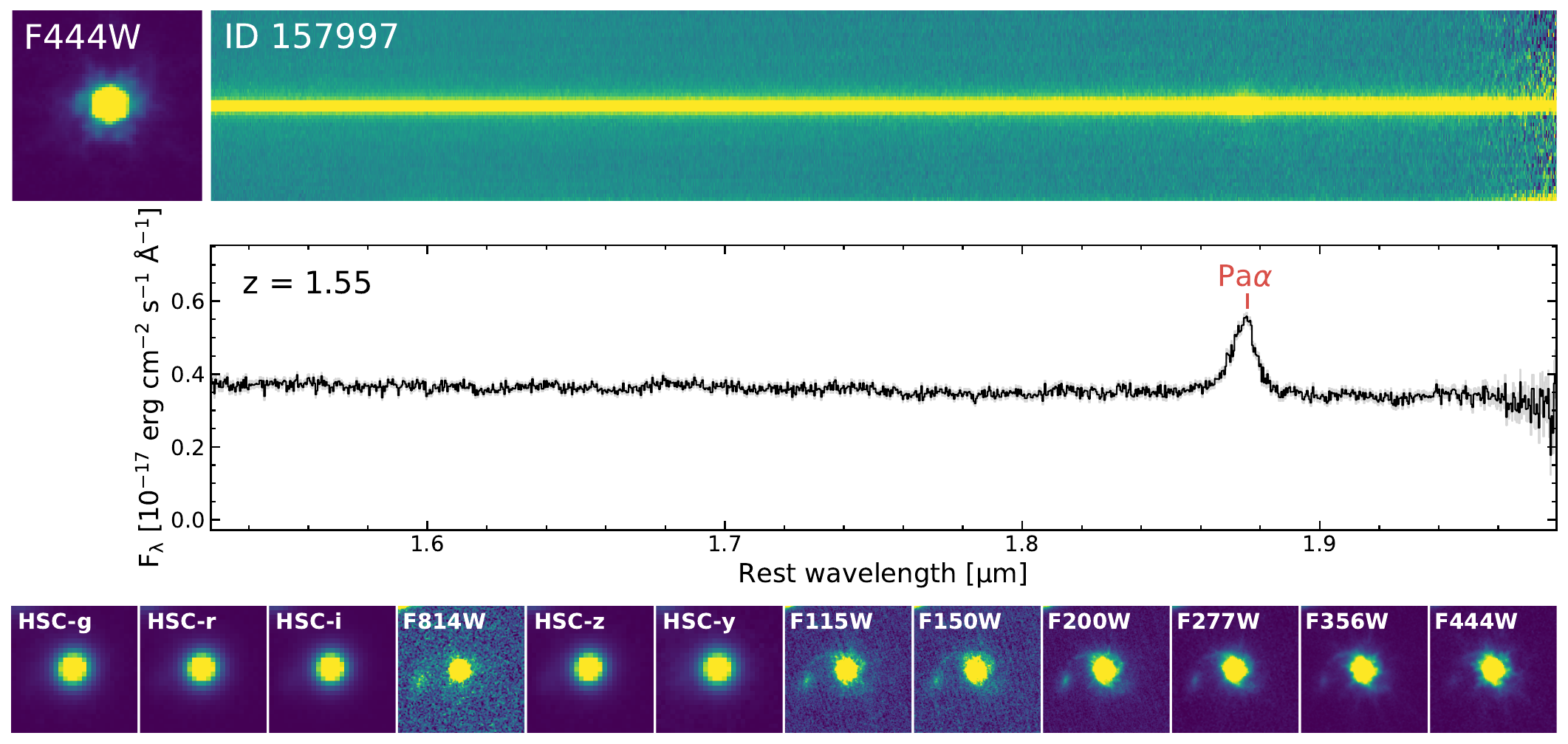}
\caption{
An example of our Pa$\alpha$ AGNs. We show the reduced 2D WFSS spectrum and the optimally extracted 1D spectrum. The broad Pa$\alpha$ line is labeled.
The size of each image is 4.5 arcsec on a side. 
The JWST F444W band image rotated to the dispersion direction is displayed in the upper left corner. Multi-band images are shown in the bottom panel, consisting of five HSC bands ($g, r, i, z, y$), one HST band (F814W), and  six JWST bands (F115W, F150W, F200W, F277W, F356W, F444W).
 \label{fig:allinfo}}
\end{figure*}

The resultant line spectra were then fitted using the double-Gaussian model as adopted above to select robust broad lines.
We required the broad Paschen line to satisfy the following criteria:
\begin{itemize}
\item[(1)] The mean SNR per pixel of the line spectrum is greater than 5;
\item[(2)] The integrated SNR of the broad component is greater than 5.
\end{itemize}
This process yielded our final broad Paschen-line AGN sample with 62 sources, consisting of 44 Pa$\alpha$ at $z=1.1-1.7$ and 18 Pa$\beta$ at $z=2.1-2.8$. The redshifts were measured by the narrow component of the double-Gaussian model. Figure \ref{fig:allinfo} shows the 2D and 1D spectra as well as multi-band images of an example from our broad Paschen-line AGN sample.
Since our sample brightness is mainly constrained by the SNR selection of broad Paschen line, the corresponding F444W magnitudes of the final broad Paschen-line AGNs are all brighter than 21 mag. This means that the 23 mag limit for the F444W photometry is sufficient to select our Paschen-line AGNs.

We verified the redshift measurements for the 62 sources with the identified broad Paschen lines. We cross-matched them with existing spectroscopic redshift catalogs. We found that 36 sources have redshifts consistent with those in the Dark Energy Spectroscopic Instrument (DESI) catalog \citep{desi2025}, and additional 18 sources are consistent with the COSMOS spectroscopic redshift compilation \citep{khostovan2025}. Another 3 sources show large discrepancies between the Paschen-line redshifts and previously reported spectroscopic redshifts. We visually inspected the archival spectra and found that their earlier line detections are less reliable than the Paschen lines. We therefore adopt the Paschen-based redshifts. For the remaining 5 sources, the Paschen-line redshifts are supported by photometric redshifts. One of them shows detections of both Pa$\alpha$ and \ion{Fe}{2}-$\lambda16440$, further confirming its redshift.

\subsection{DESI Optical Spectra} \label{desi}

To construct Paschen-based $M_{\mathrm{BH}}$ estimators, we used DESI optical spectra to obtain reference $M_{\mathrm{BH}}$ measurements. DESI has conducted a comprehensive optical spectroscopic survey of millions of quasars in the sky. The spectral resolutions of its three optical channels (blue: $3600-5900$ \AA, green: $5660-7220$ \AA, and red: $7470-9800$ \AA) are $R \sim 2700$, $R \sim 4200$, and $R \sim 4600$, respectively \citep{desicollaboration2022}. Multiple observational programs, including the survey validation, special observations, and the main survey, cover the COSMOS field \citep[e.g.,][]{scoville2007,myers2023,collaboration2024}. Therefore, the DESI quasar spectroscopy in the COSMOS field provides a deep and complete coverage. We adopted the public data products from DESI Data Release 1 \citep[DR1;][]{desi2025}. For the spectra that we used, we corrected the galactic extinction, adopting the galactic extinction curve proposed by \cite{fitzpatrick1999} with an assumed value of $R_V = 3.1$.

Within the redshift range of our broad Paschen-line AGN sample, the \ion{Mg}{2} emission line is the most reliable $M_{\mathrm{BH}}$ indicator available from the DESI spectra. We cross-matched our broad Paschen-line AGN sample with the DESI DR1 catalog to construct a calibration sub-sample with both Paschen line and \ion{Mg}{2} line coverage. To ensure reliable measurements of \ion{Mg}{2} lines, we restricted the redshift to be smaller than 2.4, which corresponds to the observed \ion{Mg}{2} wavelengths below 9500 \AA. We rebinned the matched \ion{Mg}{2} spectra by an integer number of pixels to a fixed wavelength step of 200 km s$^{-1}$ using a public code {\tt\string SpectRes} \citep{carnall2017}. We required the mean SNRs per pixel of the rebinned \ion{Mg}{2} lines to be greater than 5. The DESI-matched sample consists of 22 Pa$\alpha$ AGNs and 5 Pa$\beta$ AGNs. After estimating the dust reddening properties in Section \ref{sed}, we further refined this sample and used it to calibrate the $M_{\mathrm{BH}}$ estimators.

We also independently searched for \ion{Mg}{2} AGNs in the DESI catalog to verify the efficiency of our broad Paschen-line AGN selection. For the Pa$\alpha$ sample, we selected DESI sources with spectype==``qso", redshifts within the grism coverage, sky coordinates within the COSMOS-3D field, and the mean SNRs of \ion{Mg}{2} larger than 5. This selection yielded 25 DESI \ion{Mg}{2} AGNs. We compared them with the broad Pa$\alpha$ AGN sample. The majority of these AGNs are successfully recovered by our broad Pa$\alpha$ selection, with only three exceptions. Two sources were excluded by the compactness criterion due to their extended morphologies. One source was rejected because its broad-line components cannot satisfy our detection criteria. We performed the same analysis for the Pa$\beta$ sample and confirmed that all reliable broad Pa$\beta$ AGNs were also successfully selected. These results demonstrate that our broad Paschen-line selection efficiently recovered known DESI AGNs with reliable broad Paschen-line detections in the COSMOS-3D field.

\section{Analyses} \label{sec:analyses}

In this section, we analyze the properties of our broad Paschen-line AGNs. We fit both Paschen lines and \ion{Mg}{2} line spectra to obtain line properties for $M_{\mathrm{BH}}$. We decompose the AGNs and their host galaxies using multi-band JWST images. We also perform SED modeling to evaluate the effects of dust reddening on our AGN sample.

\subsection{Spectral Fitting} \label{spec_fit}

We perform spectral fitting for the selected Paschen lines and the corresponding \ion{Mg}{2} lines. We use a multi-Gaussian model for both Paschen lines and \ion{Mg}{2} lines. The model consists of two broad-line components with 1200 $\rm km~s^{-1}$ $<$ FWHM $<$ 12000 $\rm km~s^{-1}$ and one narrow-line component with FWHM $<$ 1200 $\rm km~s^{-1}$ \citep[e.g.,][]{shen2011a}. This multi-Gaussian model provides a flexible and reliable description of the line profiles and is commonly used in previous $M_{\mathrm{BH}}$ measurements \citep[e.g.,][]{greene2005,wang2009,pan2025b}. To minimize potential systematic biases associated with line-profile models \citep{kim2010}, we analyze all the $M_{\mathrm{BH}}$ properties based on the measurements derived from this multi-Gaussian model.

\begin{figure}[t]
\epsscale{1.2}
\plotone{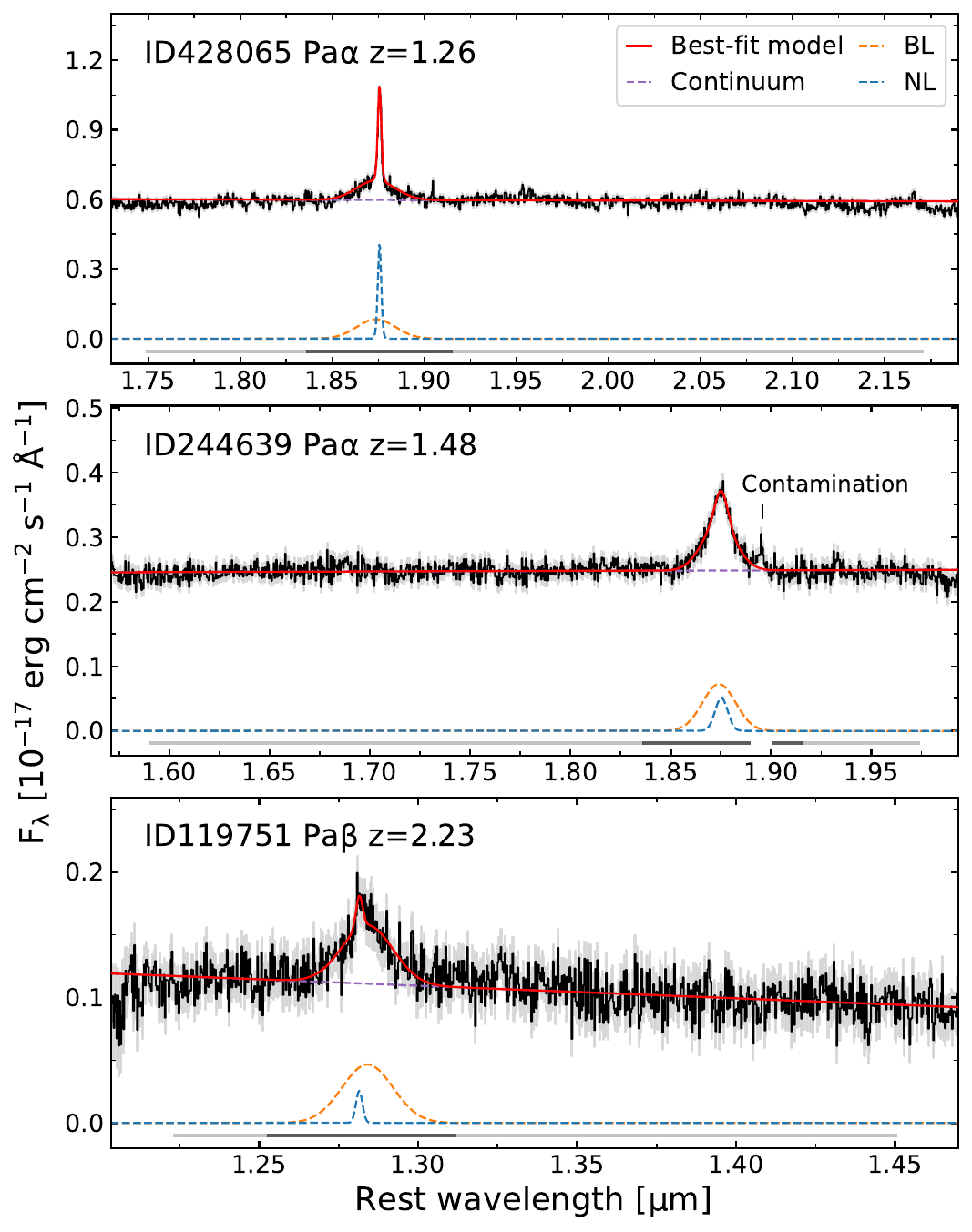}
\caption{Spectral fitting for three representative Pa$\alpha$ and Pa$\beta$ spectra. Each 1D spectrum (black) is shown along with the corresponding error spectrum (grey-shaded region). The color-coded lines show the linear continuum (purple), the broad Paschen line (BL, orange) and the narrow Paschen line (NL, blue). The red line shows the sum of all components (best-fit model). The line and continuum fitting windows are shown in dark and light grey, respectively. We label a contamination line from a nearby source in the middle panel.
 \label{fig:PaA_spec}}
\end{figure}

We fit the JWST Paschen-line spectra using the multi-Gaussian model convolved with the line-spread functions (LSFs). The construction of the NIRCam WFSS LSFs will be described in Sun et al. (in preparation.; see also \citealt{lin2025b,danhaive2025d}). The line spectra are optimally extracted, contamination-removed, and continuum-subtracted as described in Section~\ref{selection}. We apply the multi-Gaussian model using the same fitting windows and the least-square method adopted in Section~\ref{selection}. For each source, the broad-line FWHM is calculated using two broad components. We show the best-fit examples of two Pa$\alpha$ and one Pa$\beta$ spectra in Figure~\ref{fig:PaA_spec}. The Pa$\alpha$ spectrum in the middle panel has a contamination line from a nearby source, which is masked during the fitting. The redshift versus broad Paschen-line luminosity distribution of our AGN sample is shown in Figure \ref{fig:zLpa}. The measured Paschen-line luminosities and FWHMs  of the full sample are presented in Appendix~\ref{sec:alllist}.

\begin{figure}[t]
\epsscale{1.2}
\plotone{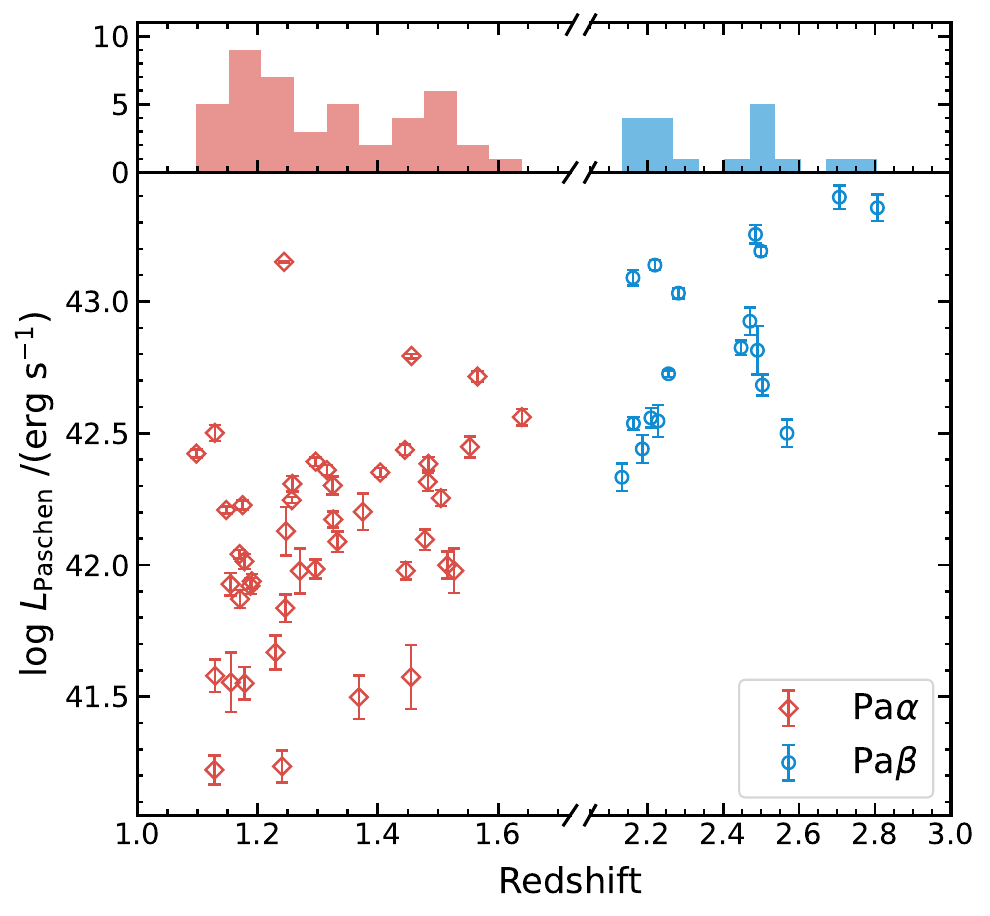}
\caption{Plot of broad Paschen-line luminosity versus redshift. The red open diamonds and blue open circles with error bars show Pa$\alpha$ and Pa$\beta$, respectively. The upper panel shows the histograms of redshift distributions.
 \label{fig:zLpa}}
\end{figure}

We carry out multi-component spectral fitting for the DESI \ion{Mg}{2} spectra. The observed continuum is decomposed into a single power-law continuum ($F_{\rm PL}$= $F_0 \lambda^{\alpha_{\lambda}}$), a Balmer continuum \citep{grandi1982}, and an iron template \citep{tsuzuki2006}. The \ion{Mg}{2} line is fitted by the multi-Gaussian model. The continuum+iron fitting windows are $\lambda_{\rm rest}=2200$--$2700$ \AA~ and $2900$--$3300$ \AA. For some spectra with shorter wavelength coverage, the fitting windows are correspondingly shorter. The fitting window for \ion{Mg}{2} is $\lambda_{\rm rest}=2700$--$2900$ \AA. 

\begin{figure}[t]
\epsscale{1.2}
\plotone{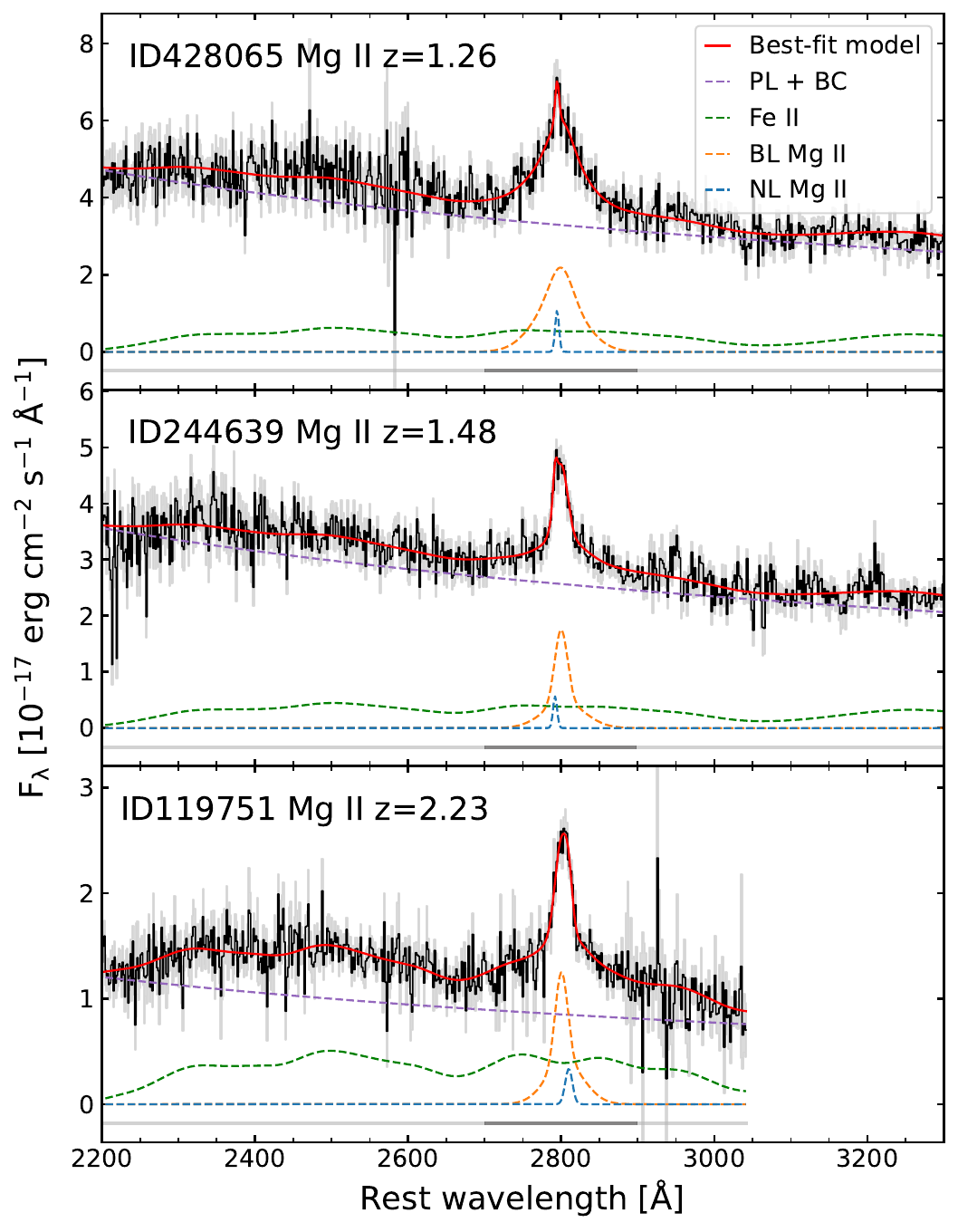}
\caption{Spectral fitting for three representative \ion{Mg}{2} spectra, corresponding to Figure \ref{fig:PaA_spec}. In each panel, the color-coded lines show the power-law continuum plus Balmer continuum (PL + BC, orange), the scaled \ion{Fe}{2} template (\ion{Fe}{2}, green), the broad \ion{Mg}{2} line (BL \ion{Mg}{2}, orange) and the narrow \ion{Mg}{2} line (NL \ion{Mg}{2}, blue). The red line shows the sum of all components (best-fit model). 
 \label{fig:MgII_spec}}
\end{figure}

The fitting procedure is the same as that described in \cite{jiang2024a}. We iteratively fit the continuum and iron template to determine the power-law normalization and slope factors, and the iron-template scale. We fit iron templates with different line widths and selected the best-fit model by minimizing $\chi^2$ over the continuum$+$iron windows. The \ion{Mg}{2} line is then fitted to the residual spectrum after subtracting the best-fit continuum$+$iron model. We also perform a 3-$\sigma$ clipping to mask out the pixels affected by strong sky lines, cosmic rays, or absorption lines. Figure \ref{fig:MgII_spec} shows best-fit results for the same sources in Figure~\ref{fig:PaA_spec}. We measure the \ion{Mg}{2} broad-line FWHMs using two broad components and luminosity $L_{3000}$ using the best-fit continuum. The \ion{Mg}{2}-based $M_\mathrm{BH}$ are then derived using the virial relation from \cite{vestergaard2009}.

For Paschen and \ion{Mg}{2} emission line fitting, we propagate the measurement uncertainties using Monte Carloc (MC) simulations. We generate 1000 mock spectra for each real spectra. The mock flux at each wavelength is the real flux with an error that is randomly drawn from a Gaussian distribution with 1 $\sigma$ equal to the observed error. We repeat our fitting processes for each version of the mock spectrum and obtained a distribution of the 1000 results for every measured parameter. The $1\sigma$ uncertainties are derived from these resulting distributions. We add additional 0.05 dex uncertainties in quadrature to $L_{3000}$ to account for the flux calibration uncertainties.

\subsection{Host Galaxy Decomposition} \label{decompse}

\begin{figure}[t]
\epsscale{1.2}
\plotone{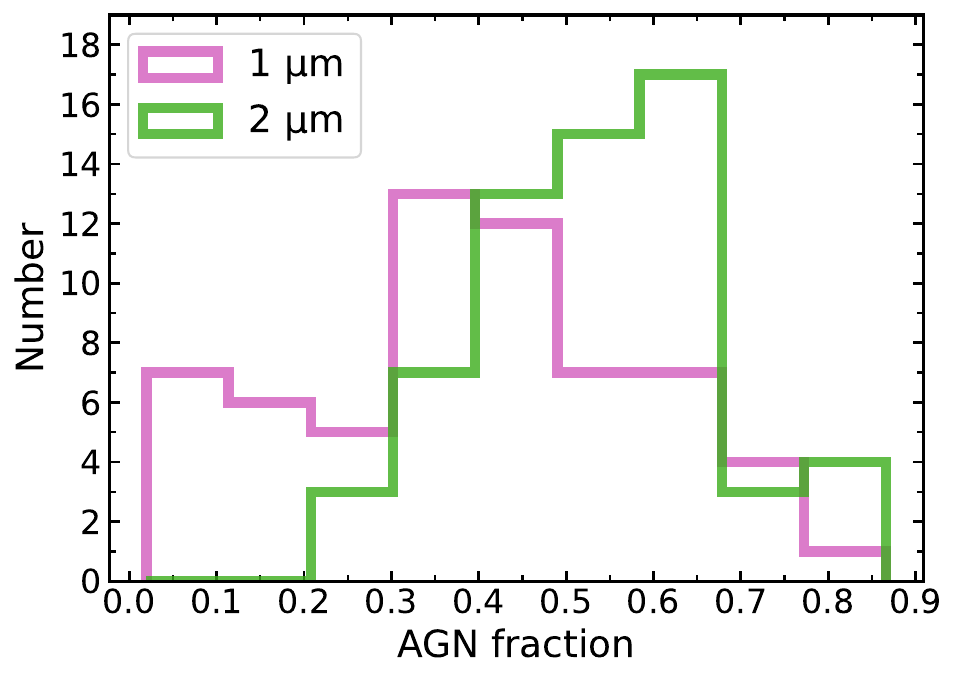}
\caption{Distribution of the measured AGN fractions. The histograms show the AGN fractions for our broad Paschen-line AGNs at the rest-frame 1 $\mathrm{\mu}$m and 2 $\mathrm{\mu}$m, indicating substantial host-galaxy contributions. 
 \label{fig:agn_fraction}}
\end{figure}

To remove host-galaxy contamination in the $M_{\mathrm{BH}}$ estimation, we decompose multi-band JWST images for our broad Paschen-line AGNs. The image decomposition is performed in six JWST bands (F115W, F150W, F200W, F277W, F356W, and F444W) using the multi-band fitting code {\tt GalfitM} \citep{haussler2022a}. In each band, the image is modeled with a PSF and a single S\'{e}rsic profile. The PSFs are constructed based on nearby point sources for each AGN using {\tt PSFEx}.

In the fitting process, we first estimate the position, morphology, and magnitude for both point source and host galaxy components using the python codes {\tt\string statmorph} \citep{statmorph} and {\tt\string photutils} \citep{larry_bradley_2024}. The derived values are fed to {\tt\string GalfitM} as initial fitting parameters. We keep the elliptical ratio, position angle, half-light radius ($R_e$), and S\'{e}rsic index constant across all bands, allow the magnitudes and positions of the PSF and S\'{e}rsic components to vary with wavelength as a fifth-order polynomial. The PSF positions are constrained to be within ±3 pixel around its ({\tt\string xpeak}, {\tt\string ypeak}), to ensure that the PSFs are fitted to the brightest peak of our target. The S\'{e}rsic positions are also limited to be within ±5 pixel around their corresponding peaks. To reduce degeneracy between PSFs and compact S\'{e}rsic models, we restrict the S\'{e}rsic index to be less than 5, and $R_e$ to be larger than 4 pixels. 

Examples of the image decomposition results are presented in Appendix~\ref{sec:show_images}. Although one PSF plus single S\'{e}rsic profile cannot fully describe the host-galaxy structure, it provides a robust separation of the PSF component. In the following analyses, we use the decomposed PSF model fluxes to represent the AGN emission. We add additional 10$\%$ uncertainties in quadrature to reflect uncertainties from the image decomposition. The host-galaxy flux is obtained by subtracting the PSF model flux from the total photometric flux. 

We then calculate AGN luminosities using the decomposed AGN flux. For each source, we interpolate the decomposed JWST-band AGN flux to measure the rest-frame 1 $\mathrm{\mu}$m and 2 $\mathrm{\mu}$m luminosities, $L_{\mathrm{1\mu m,AGN}}$ and $L_{\mathrm{2\mu m,AGN}}$. 
The measured AGN luminosities of the full sample are presented in Appendix~\ref{sec:alllist}.
The AGN fraction is defined as the ratio of the PSF flux to the total flux. For ID244208, we use the 2 $\mathrm{\mu}$m luminosity measured from the grism spectrum because its F444W image is saturated. We set the 2 $\mathrm{\mu}$m AGN fraction of ID244208 to unity. As shown in Figure~\ref{fig:agn_fraction}, the NIR luminosities of the broad Paschen-line AGNs exhibit substantial host-galaxy contributions, reinforcing the necessity of the host decomposition.

\subsection{SED Modeling} \label{sed}

We perform a multi-component SED fitting of the broad Paschen-line AGN sample to estimate their dust reddening properties. The fitting is based on the decomposed SEDs of six JWST bands (F115W, F150W, F200W, F277W, F356W, F444W) together with the total flux of one HST band (F814W) and five HSC bands ($g, r, i, z, y$). 
Our sole goal is to estimate the dust extinction that will be used to correct the reddened AGN continuum luminosity. To reduce the degeneracy associated with detailed host-galaxy properties, we model the galaxy emission by fitting galaxy templates instead of stellar population synthesis models.
We use the above multi-wavelength data to constrain a three-component SED model that consist of one reddened AGN component and two galaxy components. For the AGN component, we adopt a low-luminosity blue quasar SED model from \cite{krawczyk2013} and apply a Small Magellanic Cloud (SMC) dust-reddening law following \cite{fitzpatrick1999}:
\begin{equation}
F_{\mathrm{obs}}(\lambda)=F_{\mathrm{int}}(\lambda)\,
10^{-E(B-V)\,k(\lambda)} \times C
\end{equation}
where $F_{\mathrm{int}}(\lambda)$ is the intrinsic quasar SED, $F_{\mathrm{obs}}(\lambda)$ is the observed SED, $C$ is a normalization constant, and $k(\lambda)$ is the extinction curve, assuming $R_V$ = 3.1. The color factor $E(B-V)$ is constrained to be $0.01-2$ to stabilize the fitting. 
The host-galaxy emission is modeled using one elliptical and one spiral galaxy SED template, following the literature \cite[e.g.,][]{assef2010,kim2010}. The galaxy template is individually scaled to fit the data.

The SED fitting is carried out by iteratively fitting the AGN and host-galaxy components. We first fit the galaxy model to the image-decomposed JWST galaxy fluxes, and then fit the AGN model. Because the HST and HSC images are not decomposed, we calculate the AGN fluxes by subtracting the best-fit galaxy model from the total fluxes. The AGN model is fitted to the galaxy-subtracted fluxes in the HST and HSC bands and the image-decomposed AGN fluxes in the JWST bands. To refine the galaxy model, we calculate the galaxy fluxes in the HST and HSC bands by subtracting the best-fit AGN model from the total fluxes. We then refit the galaxy model to the AGN-subtracted fluxes in the HST and HSC bands and the image-decomposed galaxy fluxes in the JWST bands. The AGN model is subsequently updated. Each fitting is performed using the least-square method. We repeat this procedure until all free parameters converge. 

We account for both measured and systematic uncertainties in the SED fitting. We add 30$\%$ uncertainties in quadrature to the total flux to account for the intrinsic scatter of the SED models. The $\chi^2$ values of the best-fit models are therefore distributed around unity. All uncertainties are propagated to the derived parameters using MC simulations similar to those described in Section~\ref{spec_fit}. 

We obtain the $E(B-V)$ values and rest-frame UV luminosities of the AGN component from the SED fitting. 
The $E(B-V)$ values of the full sample are presented in Appendix~\ref{sec:alllist}. To calibrate the \ion{Mg}{2}-based $M_{\mathrm{BH}}$ estimators, we require that AGNs in the calibration sample have negligible dust attenuation with $E(B-V)<0.2$. As a result, three Pa$\alpha$ AGNs (ID770023, ID205967 and ID7741) are excluded from the DESI-matched sample. The final calibration sample consists of 19 Pa$\alpha$ AGNs and 5 Pa$\beta$ AGNs. The measured UV luminosities $L_{3000}$ of the calibration sample reliably trace the intrinsic AGN UV emission. Their modeled AGN fractions at rest-frame 3000~\AA\ all exceed 80$\%$ with a median value of 93$\%$, indicating minimal host-galaxy contamination. Other broad Paschen-line AGNs exhibit varying degrees of dust attenuation. We correct their intrinsic AGN luminosities using the derived $E(B-V)$ values in the following analyses.

\section{Results} \label{sec:results}

\subsection{Correlations Between Line Properties} \label{subsec:correlations}

\begin{figure*}[t]
\epsscale{1.}
\plotone{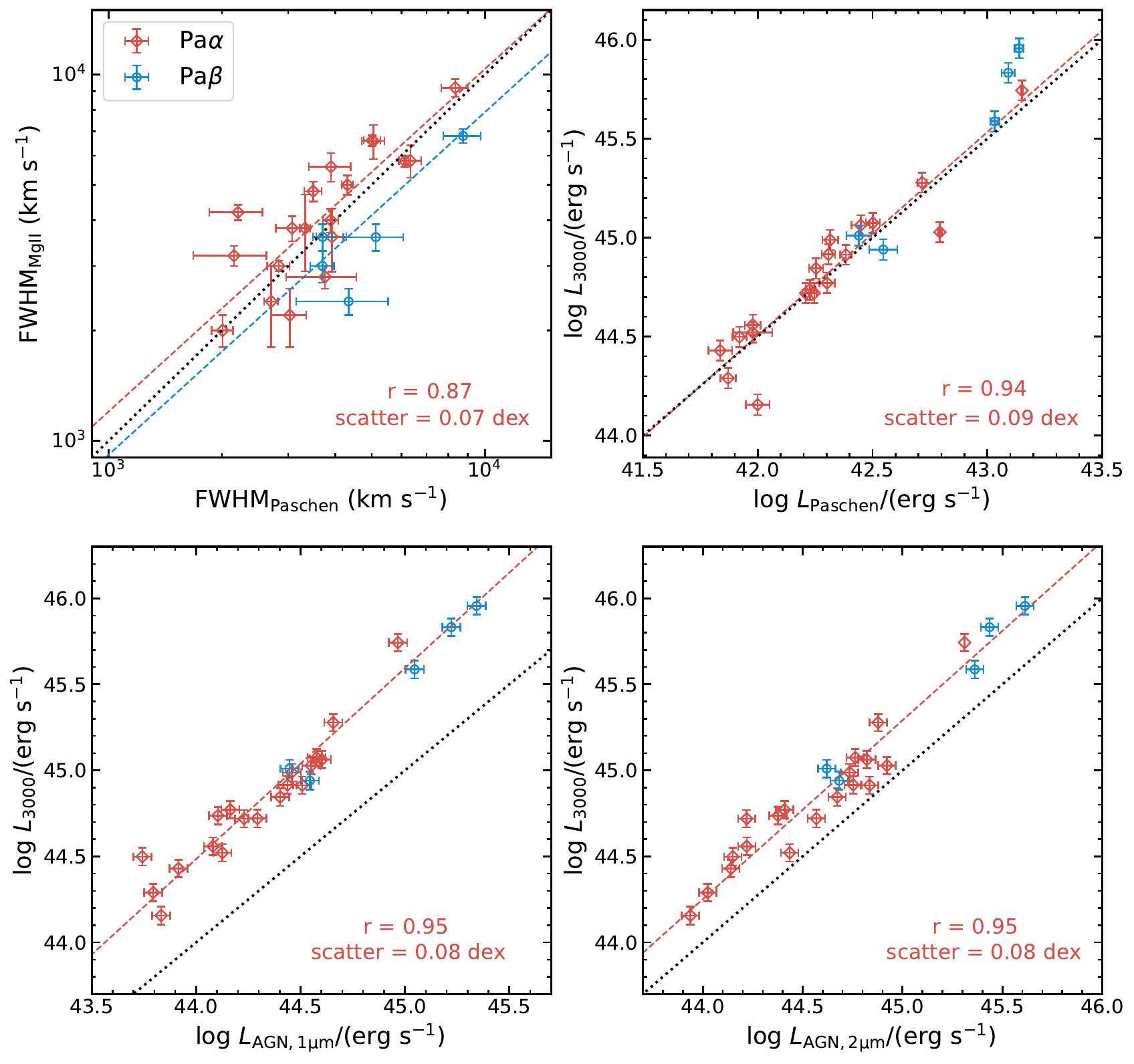}
\caption{Comparisons between the properties of UV and NIR indicators. The red open diamonds and blue open circles with error bars show the properties of Pa$\alpha$ and Pa$\beta$, respectively. The red and blue dashed lines show the best-fit correlations for Pa$\alpha$ and Pa$\beta$, respectively. 
The black dotted line indicates the 1:1 relation, except in the upper-right panel, where it is shifted upward by 2.5 dex.
The upper-left panel compares the FWHMs between \ion{Mg}{2} and Paschen lines. Other panels compare the UV luminosities with NIR line and AGN continuum luminosities.
 \label{fig:correla}}
\end{figure*}

In this subsection, we present empirical correlations between the properties of \ion{Mg}{2} line and the Paschen lines.  We also examine correlations between the UV and NIR continuum luminosities. These correlations enable the calibration of $M_{\mathrm{BH}}$ estimators from \ion{Mg}{2} to the Paschen lines. For each correlation, we perform linear-model fitting using the Orthogonal Distance Regression (ODR) method, which accounts for measurement uncertainties in both axes. The equations are
\begin{equation}
\mathrm{log}\mathrm{(\frac{FWHM_{\text{Mg\,\textsc{ii}}}}{10^3\ km\ s^{-1}})} = X_1 \times \mathrm{log(\frac{FWHM}{10^3\ km\ s^{-1}})} + Y_1, 
\end{equation}
\begin{equation}
\mathrm{log(\frac{\textit{L}_{3000}}{10^{44}\ erg\ s^{-1}})} = X_2\times \mathrm{log(\frac{\textit{L}}{10^{44}\ erg\ s^{-1}})} + Y_2,
\end{equation}
where $X_1$ and $Y_1$ are the correlation coefficients for FWHMs, $X_2$ and $Y_2$ are the correlation coefficients for luminosities (Table~\ref{tab:corr}). 

The FWHMs of the \ion{Mg}{2} and Pa$\alpha$ are tightly correlated, with a Pearson correlation coefficient of $r = 0.87$ and a scatter of 0.07 dex. The Pa$\alpha$ FWHMs are slightly smaller than those of \ion{Mg}{2} by $\sim 500$ km s$^{-1}$, suggesting that Pa$\alpha$ originates from a slightly outer region of the BLR compared to \ion{Mg}{2}. Although only five Pa$\beta$ sources are available, they exhibit systematically larger FWHMs than Pa$\alpha$. These trends are consistent with previous studies \citep{kim2010,pan2025b}. 
Due to the limited sample size, we fix the slope and scale the FWHM correlation of Pa$\alpha$ to obtain that for Pa$\beta$. The resulting FWHM correlations are shown in the upper-left panel of Figure~\ref{fig:correla}. The best-fit coefficients are listed in Table~\ref{tab:corr}.

\begin{deluxetable}{ccc}[t]
\tablecaption{Coefficients of the Empirical Correlations \label{tab:corr}}
\setlength{\tabcolsep}{15pt}
\tablehead{
\colhead{FWHM}  & \colhead{$X_1$} & \colhead{$Y_1$}
}
\startdata
Pa$\alpha$ & 0.93 $\pm$ 0.11 & 0.08 $\pm$ 0.08 \\
Pa$\beta$ & 0.93 $\pm$ 0.11 & $-$0.04 $\pm$ 0.03 \\
\hline
\hline
\colhead{$L$}  & \colhead{$X_2$} & \colhead{$Y_2$} \\
\hline
Pa$\alpha$ & 1.03 $\pm$ 0.08 & 2.56 $\pm$ 0.14 \\
AGN, 1$\mu$m & 1.11 $\pm$ 0.09 & 0.48 $\pm$ 0.04 \\
AGN, 2$\mu$m & 1.04 $\pm$ 0.08 & 0.25 $\pm$ 0.05 \\
\enddata
\end{deluxetable}

We also investigate the correlations between the UV luminosity $L_{3000}$ and the NIR luminosities, namely broad-line luminosity $L_{\mathrm{Pa\alpha}}$, NIR AGN luminosity $L_{\mathrm{AGN,1\mu m}}$, and $L_{\mathrm{AGN,2\mu m}}$.
All the NIR luminosities are strongly correlated with $L_{3000}$, with $r \sim 0.95$ and scatters smaller than 0.1. Because the Pa$\beta$ luminosity distribution does not differ significantly from that of Pa$\alpha$, we do not derive a separate luminosity relation for Pa$\beta$. The results of luminosity correlations are shown in Figure \ref{fig:correla}. The best-fit coefficients are listed in Table~\ref{tab:corr}. While the NIR continuum luminosities include significant contributions from the dusty torus, these tight correlations indicate that they can serve as reliable proxies for $M_{\mathrm{BH}}$ \citep[e.g.,][]{netzer2015}.

\subsection{BH Mass Estimators} \label{subsec:mbh_estimators}

We derive new $M_{\mathrm{BH}}$ estimators based on broad Paschen emission lines using the calibration sample. We construct three types of mass estimators, in which the $R_{\mathrm{BLR}}$ is inferred from the Paschen-line luminosity $L_{\mathrm{Paschen}}$, the AGN continuum luminosity $L_{\mathrm{AGN,1\mu m}}$, and $L_{\mathrm{AGN,2\mu m}}$, respectively. The velocity dispersion term is provided by the FWHMs of the Paschen lines. We adopt $M_{\mathrm{BH}}$  derived from the \ion{Mg}{2} lines as reference values. We perform a multivariable linear regression fit using the ODR method with the following equation:
\begin{equation}
\begin{split}
\mathrm{log(\frac{\textit{M}}{\textit{M}_{\odot}}}) & = A\times \mathrm{log(\frac{FWHM}{10^3\ km\ s^{-1}})} \\
&+ B\times\mathrm{log(\frac{\textit{L}}{10^{44}\ erg\ s^{-1}})} +C.
\end{split}
\end{equation}
Under the assumptions of virial theorem and the empirical $R$–$L$ relation, the expected values of the coefficients are $A = 2$ and $B = 0.5$.

\begin{figure*}[t]
\epsscale{1.2}
\plotone{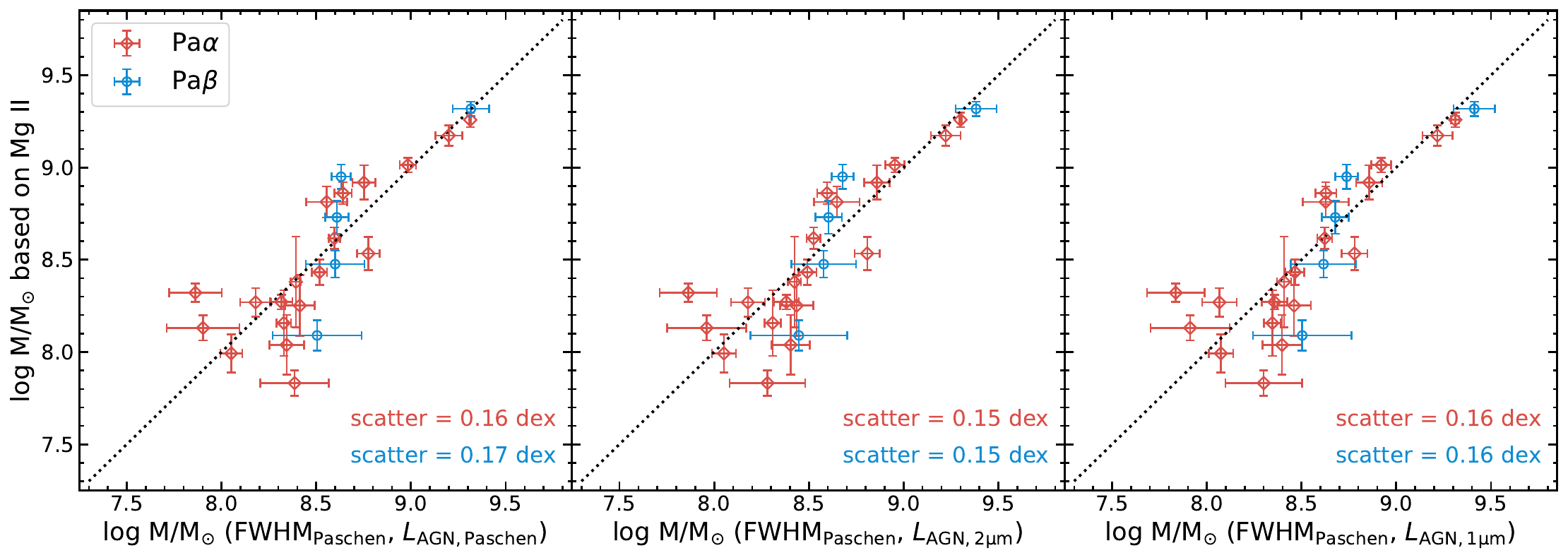}
\caption{Comparisons between the \ion{Mg}{2}-based $M_{\mathrm{BH}}$ and the Paschen-based $M_{\mathrm{BH}}$ values. The red open diamonds and blue open circles with error bars show $M_{\mathrm{BH}}$ based on Pa$\alpha$ and Pa$\beta$, respectively. The black dotted line in each panel indicates the 1:1 relation.
 \label{fig:mbh_cal}}
\end{figure*}

We first perform the fitting with three free parameters. The sample size of Pa$\beta$ is small, so we primarily use the Pa$\alpha$ sample. The resultant coefficients ($A$, $B$, $C$) for the three luminosities ($L_{\mathrm{Pa\alpha}}$, $L_{\mathrm{AGN,2\mu m}}$, and $L_{\mathrm{AGN,1\mu m}}$) are summarized in Table~\ref{tab:factor_mbh}.
The Pa$\beta$-based estimators are then obtained by scaling those of Pa$\alpha$ using the FWHM correlations, which only affects the constant term: $C_{\mathrm{Pa}\beta}=C-0.126\times A$. Figure~\ref{fig:mbh_cal} compares the $M_{\mathrm{BH}}$ values derived from our three estimators with those based on the \ion{Mg}{2} line. The intrinsic calibration scatters are $\sim$0.16 dex. The derived coefficients are consistent with the expected values within the $1\sigma$ uncertainties. Given that the coefficient $A$ is close to the virial expectation of 2, we additionally present fittings with $A$ fixed to 2. 
These fixed-slope fittings yield similar results with reduced uncertainties, which are also listed in Table~\ref{tab:factor_mbh}.

\begin{deluxetable*}{ccccccc}[t]
\tablecaption{Factors of $M_{\mathrm{BH}}$ Estimators\label{tab:factor_mbh}}
\setlength{\tabcolsep}{3.5pt}
\tablehead{
\colhead{Parameter} &
\multicolumn{2}{c}{$L_{\mathrm{Pa\alpha}}$} &
\multicolumn{2}{c}{$L_{\mathrm{AGN,1\mu m}}$} &
\multicolumn{2}{c}{$L_{\mathrm{AGN,2\mu m}}$} \\
\cline{2-3}\cline{4-5}\cline{6-7}
& \colhead{fixed $A=2$} & \colhead{all free}
& \colhead{fixed $A=2$} & \colhead{all free}
& \colhead{fixed $A=2$} & \colhead{all free}
}
\startdata
$A$ & 2 & 1.96 $\pm$ 0.27 & 2 & 2.19 $\pm$ 0.26 & 2 & 2.16 $\pm$ 0.26 \\
$B$ & 0.44  $\pm$ 0.09 & 0.44 $\pm$ 0.10 & 0.50 $\pm$ 0.10 & 0.48 $\pm$ 0.11 & 0.43 $\pm$ 0.08 & 0.41 $\pm$ 0.09 \\
$C$ & 8.11  $\pm$ 0.13 & 8.15 $\pm$ 0.28 & 7.24 $\pm$ 0.06 & 7.13 $\pm$ 0.17 & 7.15 $\pm$ 0.07 & 7.06 $\pm$ 0.16
\enddata
\tablecomments{We recommend the $M_{\mathrm{BH}}$ estimators with fixed $A=2$, as the best-fit value of $A$ is consistent with the virial expectation of 2 (see also Section \ref{subsec:mbh_estimators}).}
\end{deluxetable*}

We evaluate the impact of host-galaxy contamination on the continuum-based $M_{\mathrm{BH}}$ estimators. We perform the fitting using the Pa$\alpha$ sample with total continuum luminosity $L_{\mathrm{1\mu m}}$ and $L_{\mathrm{2\mu m}}$.  The resultant coefficients ($A$, $B$, $C$) are ($2.05 \pm 0.25,0.70 \pm 0.14,6.80 \pm 0.18$) for $L_{\mathrm{1\mu m}}$ and ($2.57 \pm 0.28,0.76 \pm 0.16,6.40 \pm 0.24$) for $L_{\mathrm{2\mu m}}$. In both cases, the coefficient $B$ deviates significantly from the expected value of 0.5. The coefficient $A$ for $L_{\mathrm{2\mu m}}$ also deviates from 2. These deviations indicate that host-galaxy contamination can bias continuum-based $M_{\mathrm{BH}}$ estimators, when the host contribution is not properly removed.

\begin{figure}[t]
\epsscale{1.1}
\plotone{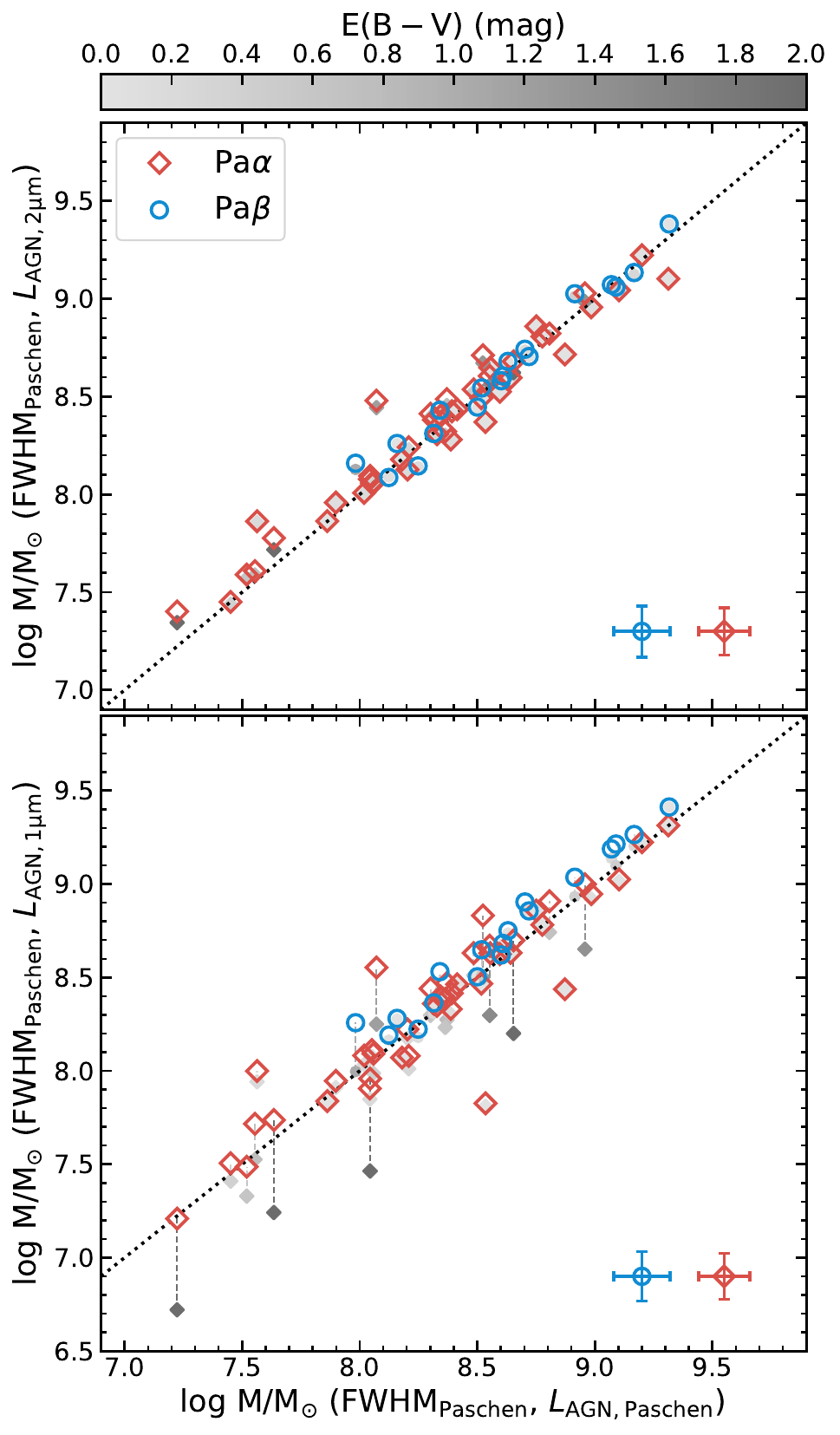}
\caption{Comparisons of the Paschen-based $M_{\mathrm{BH}}$ from derived from three estimators. 
The red open diamonds and blue open circles show dust-corrected $M_{\mathrm{BH}}$ based on Pa$\alpha$ and Pa$\beta$, respectively. Typical measurement uncertainties are shown in the lower-right corner of each panel. Grey data points, color-coded by $E(B-V)$, show $M_{\mathrm{BH}}$ before dust correction and are connected to their corresponding corrected values with grey dashed lines. The black dotted line in each panel indicates the 1:1 relation. The dust-corrected $M_{\mathrm{BH}}$ measurements are well consistent.
 \label{fig:mbh_all}}
\end{figure}

We apply our $M_{\mathrm{BH}}$ estimators to the full sample of broad Paschen-line AGNs. Since dust attenuation may reduce the observed AGN continuum when $E(B-V)$ is non-negligible, continuum-based estimators require correction to recover the intrinsic AGN luminosity. Figure \ref{fig:mbh_all} compares $M_{\mathrm{BH}}$ derived from three estimators. Results obtained without dust correction are shown as grey symbols. Dust-corrected values are indicated by red and blue markers. The $L_{\mathrm{AGN,2\mu m}}$-based $M_{\mathrm{BH}}$ is minimally affected by dust attenuation and consistent with the $L_{\mathrm{Paschen}}$-based estimates. Prior to dust correction, the $L_{\mathrm{AGN,1\mu m}}$-based $M_{\mathrm{BH}}$ show larger deviations for sources with higher $E(B-V)$. The dust-corrected $L_{\mathrm{AGN,1\mu m}}$-based $M_{\mathrm{BH}}$ become consistent with the Paschen-based values. These results indicate that $M_{\mathrm{BH}}$ estimators based on the 1 $\mu$m continuum should be applied with caution, as significant dust attenuation needs to be corrected.

To summarize, we recommend the following three $M_{\mathrm{BH}}$ estimators based on Pa$\alpha$ with a fixed virial coefficient of $A=2$:
\begin{equation}
\begin{split}
\mathrm{log(\frac{\textit{M}}{\textit{M}_{\odot}}}) & = 2\times \mathrm{log(\frac{FWHM_{Pa\alpha}}{10^3\ km\ s^{-1}})} \\
&+ 0.44\times\mathrm{log(\frac{\textit{L}_{Pa\alpha}}{10^{44}\ erg\ s^{-1}})} + 8.11.
\end{split}
\end{equation}
\begin{equation}
\begin{split}
\mathrm{log(\frac{\textit{M}}{\textit{M}_{\odot}}}) & = 2\times \mathrm{log(\frac{FWHM_{Pa\alpha}}{10^3\ km\ s^{-1}})} \\
&+ 0.50\times\mathrm{log(\frac{\textit{L}_{\mathrm{AGN,1\mu m}}}{10^{44}\ erg\ s^{-1}})} + 7.24.
\end{split}
\end{equation}
\begin{equation}
\begin{split}
\mathrm{log(\frac{\textit{M}}{\textit{M}_{\odot}}}) & = 2\times \mathrm{log(\frac{FWHM_{Pa\alpha}}{10^3\ km\ s^{-1}})} \\
&+ 0.43\times\mathrm{log(\frac{\textit{L}_{\mathrm{AGN,2\mu m}}}{10^{44}\ erg\ s^{-1}})} + 7.15.
\end{split}
\end{equation}
We do not fix the coefficient $B$, since the slope of the R–L relation could slightly deviate from 0.5 across different wavelength ranges \citep[e.g.,][]{Mandal2024,Amorim2024}. 
The free-parameter $M_{\mathrm{BH}}$ estimators can also be found in Table~\ref{tab:factor_mbh}. The Pa$\beta$-based estimators can be derived by adjusting the coefficient $C$ as described earlier.

\subsection{Luminosity Functions} \label{subsec:LF}

We build AGN LFs using our broad Paschen-line AGN sample for Pa$\alpha$ and Pa$\beta$ AGNs at redshift $z\sim 1.4$ and $z\sim 2.4$, respectively. We perform simulations to estimate our sample completeness by computing a selection function, defined as the probability that an AGN with a given broad-line luminosity $L$ at a given redshift $z$ can be selected by our selection procedure. The incompleteness arises from the grism wavelength coverage and the broad-line detection limits.

In the simulation, we randomly generate 1000 sky positions within the COSMOS-3D WFSS field and extract the corresponding 1D spectra. This step accounts for the background flux and wavelength coverage at different positions. To mimic real spectra, we construct a model that consists of a linear continuum and an emission line composed of one broad Gaussian plus one narrow Gaussian component. The continuum slope, line centroid, and Gaussian parameters are randomly drawn from the observed parameter space. The continuum normalization is correlated with the broad-line flux. We estimate this correlation from the observed data and apply it in the simulations, with a random scatter added to reproduce the intrinsic dispersion. For each extracted spectrum, we insert the spectral model for 200 times to obtain 200 mock spectra.  We then apply the same line detection and broad-line selection procedures that we use for the real data. We remove sources whose recovered FWHM or flux deviate significantly from the input values to avoid fake detections. We perform these simulations separately for Pa$\alpha$ and Pa$\beta$. The resultant selection functions are computed as the fraction of simulated lines in each ($L$, $z$) bin that satisfy our selection criteria.

\begin{figure*}[t]
\epsscale{1.1}
\plotone{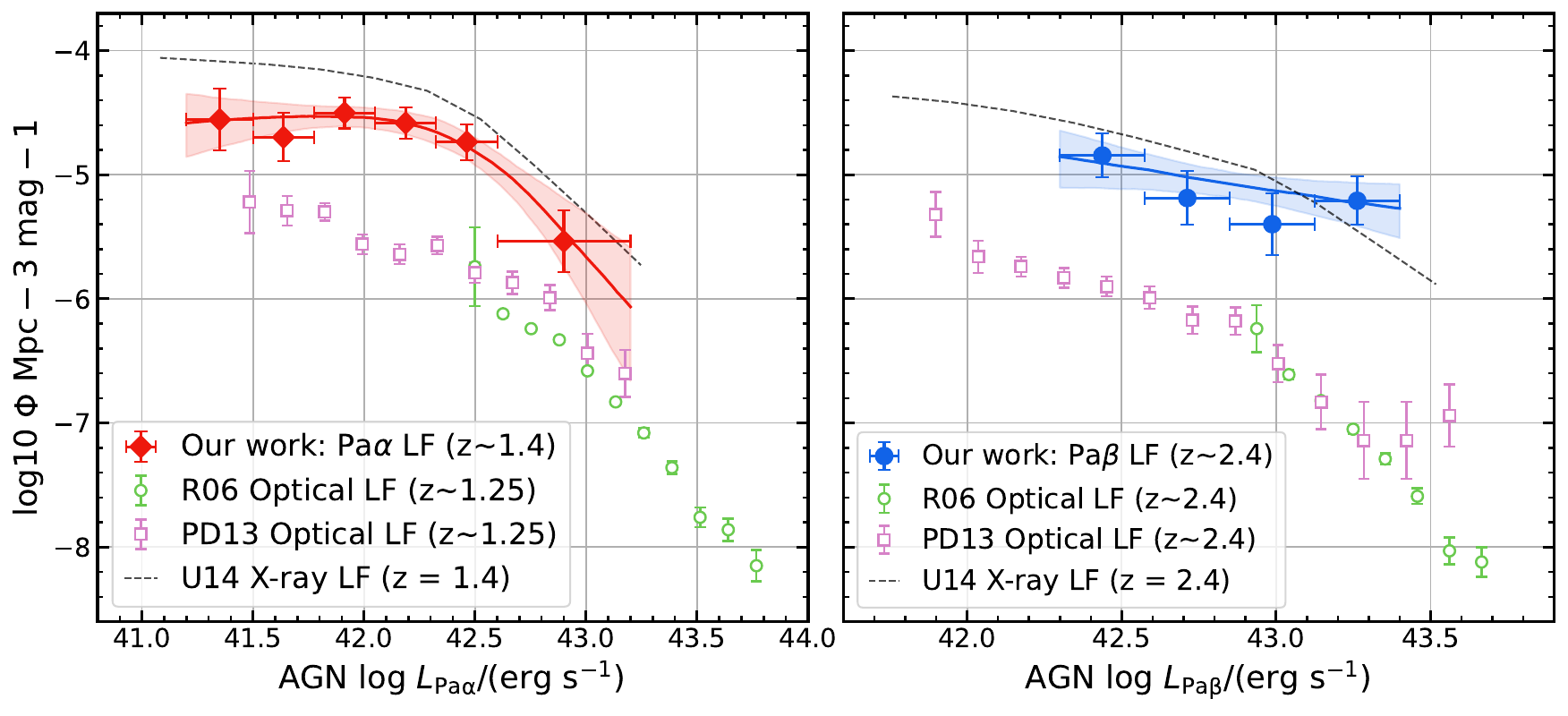}
\caption{AGN LFs of Pa$\alpha$ at $z\sim 1.4$ (left) and Pa$\beta$ at $z\sim 2.4$ (right). The red diamonds and blue circles represent our binned LFs of Pa$\alpha$ and Pa$\beta$, respectively. Its horizontal bars indicate the luminosity ranges covered by the bins, and the vertical bars show the 1$\sigma$ errors of the LFs. The corresponding red and blue lines with the shaded regions represent our best-fit LFs with their 1$\sigma$ regions. For both Pa$\alpha$ and Pa$\beta$ LFs, we compare our results with converted LFs from the literature. The green open circles and purple open rectangles show optical LFs from R06 \citep{richards2006c} and PD13 \citep{palanque-delabrouille2013}, respectively. The black dashed lines are X-ray LFs calculated following U14 \citep{ueda2014}.
 \label{fig:lfs}}
\end{figure*}

We first compute the binned LF for Pa$\alpha$ and Pa$\beta$ AGNs using the traditional 1/$V_a$ method \citep{avni1980}, accounting for their selection functions. The uncertainty of the binned LF is calculated by the Poisson error. The mean cosmic variance in COSMOS-3D field is $10-20\%$ for galaxies with $M_{\mathrm{stellar}} \sim 10^{10} M_{\odot}$ at $z\sim 1-3$ \citep{moster2011}, much smaller than the statistic errors here, so we ignore it. The derived binned LFs are listed in Table \ref{tab:binned_LF} and shown in Figure \ref{fig:lfs}. We notice an excess in the brighter end of the Pa$\beta$ LF, corresponding to an overdensity of the known proto-supercluster ``Hyperion'' in the COSMOS field at $z\sim 2.45$ \citep{cucciati2018,forrest2025}.

\begin{deluxetable}{ccccc}[t]
\tablecaption{Binned Paschen-line Luminosity Function\label{tab:binned_LF}}
\setlength{\tabcolsep}{10pt}
\tablehead{
\colhead{$\log L_{\mathrm{P}\alpha}$} &
\colhead{$\Delta \log L$} &
\colhead{$\log \Phi^{a}$} &
\colhead{$N$} &
\colhead{$N_{corr}^{b}$}
}
\startdata
\multicolumn{4}{c}{$\mathrm{Pa}\alpha \quad z\sim1.4$} \\
\hline
41.35 & 0.300 & $-4.56\pm$0.25 & 3 & 13 \\
41.64 & 0.275 & $-4.70\pm$0.19 & 5 & 10 \\
41.91 & 0.275 & $-4.50\pm$0.13 & 12 & 14 \\
42.19 & 0.275 & $-4.58\pm$0.13 & 12 & 13 \\
42.46 & 0.275 & $-4.74\pm$0.14 & 9 & 11 \\
42.90 & 0.600 & $-5.54\pm$0.25 & 3 & 3 \\
\hline
\multicolumn{4}{c}{$\mathrm{Pa}\beta \quad z\sim2.4$} \\
\hline
42.44 & 0.275 & $-4.84\pm$0.18 & 6 & 9 \\
42.71 & 0.275 & $-5.19\pm$0.22 & 4 & 5 \\
42.99 & 0.275 & $-5.40\pm$0.25 & 3 & 3 \\
43.26 & 0.275 & $-5.21\pm$0.19 & 5 & 6 \\
\enddata
\tablenotetext{a}{Luminosity function in units of $\mathrm{Mpc^{-3}\,dex^{-1}}$.}
\tablenotetext{b}{Number of Paschen-line AGNs in each bin after completeness correction.}
\end{deluxetable}

We then calculate the parametrized LFs. For Pa$\alpha$ AGNs, we adopt a double power-law (DPL) function:
\begin{equation}
\Phi(L) = \frac{\Phi^*}{10^{\gamma_1(L-L^*)} + 10^{\gamma_2(L-L^*)}},
\end{equation}
where $\Phi^*$ represents the number density normalization, $\gamma_1$ and $\gamma_2$ are respectively the faint-end and bright-end slopes, and $L^*$ is the break luminosity. For Pa$\beta$ AGNs, given the smaller sample size and limited luminosity coverage, we fit the LF with a single power-law (SPL) model:
\begin{equation}
\Phi(L) = \frac{\Phi^*}{10^{\gamma_1(L-L^*)}},
\end{equation}
where log $L^*$ is fixed to 42.5, $\Phi^*$ is the number density normalization and $\gamma_1$ is the slope. We fit the LFs using the maximum likelihood method \citep{marshall1983}. We perform a Markov Chain Monte Carlo (MCMC) analysis with a python code {\tt\string emcee} \citep{foreman-mackey2013} to determine the best fit of the free parameters. The best-fit LFs are shown in Figure \ref{fig:lfs} with the corresponding parameters listed in Table \ref{tab:LF}. We compare our Pa$\alpha$ and Pa$\beta$ LFs with previous AGN LFs in Section \ref{subsec:comparison_lf}.

\begin{deluxetable}{ccccc}[t]
\tablecaption{Parametrized Paschen-line Luminosity Function\label{tab:LF}}
\setlength{\tabcolsep}{4pt}
\tablehead{ 
\colhead{$z$} &
\colhead{$\log L^*$} &
\colhead{$\log \Phi^{*}$} &
\colhead{$\gamma_1$} &
\colhead{$\gamma_2$} 
}
\startdata
1.1$-$1.7 & $42.43_{-0.24}^{+0.23}$ & $-4.41_{-0.26}^{+0.17}$ & $-0.14_{-0.40}^{+0.36}$ & $2.21_{-0.87}^{+1.10}$ \\
2.1$-$2.8 & 42.5 & $-4.73_{-0.33}^{+0.30}$ & $0.39_{-0.35}^{+0.35}$ & $--$ \\
\enddata
\end{deluxetable}

\section{Discussion} \label{sec:discussion}

\subsection{Comparison with Previous $M_{\mathrm{BH}}$ Estimators}

We compare our $M_{\mathrm{BH}}$ estimators with those in the literature \citep{kim2010,landt2013}, by applying their relations to our broad Paschen-line AGN sample. \cite{kim2010} provided estimators based on Pa$\alpha$ and Pa$\beta$ line luminosities. Our line-based $M_{\mathrm{BH}}$ estimates show typical differences of 0.10 dex with 0.05 dex scatter for Pa$\alpha$ and 0.07 dex with 0.10 dex scatter for Pa$\beta$ relative to their calibrations. \cite{landt2013} derive Paschen-line $M_{\mathrm{BH}}$ from continuum luminosities, accounting for host-galaxy contributions using HST imaging. We adopt dust-corrected AGN luminosities at $1\,\mu\mathrm{m}$ and obtain the typical difference of 0.06 dex with 0.12 dex scatter. 
Overall, our $M_{\mathrm{BH}}$ estimators are consistent with previous calibrations within our intrinsic calibration scatters.

\subsection{Comparison with Previous AGN Luminosity Functions} \label{subsec:comparison_lf}

Our NIR broad Paschen-line selection using JWST grism data reduces the incompleteness caused by dust attenuation and photometric pre-selection. Therefore, the broad Paschen-line AGN sample is expected to provide a more comprehensive census than the previous optically selected samples. To evaluate this, we convert optical and X-ray LFs at similar redshifts into the broad Paschen-line LFs and compare them with our LFs.

The optical LFs are adopted from \cite{richards2006c} and \cite{palanque-delabrouille2013}. They are constructed from optically selected and spectroscopically confirmed broad-line AGN samples. Their luminosity indicators are the absolute magnitudes $M_i(z=2)$ and $M_g(z=2)$, respectively. We convert the latter to $M_i(z=2)$ following \cite{ross2013a}. 
We compute the $M_i(z=2)$ of our broad Paschen-line AGNs using the dust-corrected AGN SED model derived in Section~\ref{sed}. We then quantify the relation between $M_i(z=2)$ and $L_{\mathrm{Paschen}}$ by fitting a linear equation log$(L_{\mathrm{Paschen}}) = a \times M_i(z=2) + b$ to our sample. The best-fit parameters ($a,b$) are $(-0.42 \pm 0.02,31.9 \pm 0.58)$ for Pa$\alpha$ AGNs and $(-0.35 \pm 0.03,33.6 \pm 0.78)$ for Pa$\beta$ AGNs. Using these relations, we convert the optical LFs into the Paschen-line luminosity frame.

The parametrized X-ray LFs are adopted from \cite{ueda2014}, who use the rest-frame $2$--$10$~keV luminosity $L_{\mathrm{X, 2-10keV}}$ as the luminosity indicator. We cross-match our broad Paschen-line AGNs with the Chandra COSMOS Legacy Survey catalog \citep{civano2016} and obtain 52 AGNs with X-ray detections. We calculate their $L_{\mathrm{X, 2-10keV}}$ and fit the relation log$(L_{\mathrm{Paschen}}) = c \times L_{\mathrm{X, 2-10keV}} + d$. We do not separate Pa$\alpha$ and Pa$\beta$ AGNs, as they follow similar correlations. The best-fit parameters are $c = 1.35 \pm 0.06$ and $d = -17.39 \pm 2.79$. The X-ray LFs are then converted to the Paschen-line luminosity frame. 

The converted optical and X-ray LFs are shown in Figure~\ref{fig:lfs}. At $z\sim1.4$, the Pa$\alpha$ AGN LF is 3--5 times higher than the optical LFs, indicating a more complete and effective identification of broad-line AGNs. 
Our Pa$\alpha$ AGN LF is 1--3 times lower than the X-ray LF. On one hand, our broad-line selection only traces Type 1 AGNs, whereas the X-ray LF also includes narrow-line Type 2 AGNs. On the other hand, this underestimate may also partly arise from the contribution of Seyfert galaxies. Due to the limitation of grism spectroscopy, we cannot reliably distinguish AGN broad-line emission from extended host-galaxy emission. Sources with strong extended host galaxies are excluded from our analyses, which could be included in the X-ray LF.
The complete AGN population could be even larger if we include middle-infrared selected AGNs \citep{lyu2022a,lyu2024}.
At $z\sim2.4$, the faint end of the Pa$\beta$ AGN LF shows a similar trend. The bright end of the Pa$\beta$ LF exceeds the X-ray LF, likely due to the presence of the proto-supercluster mentioned in Section~\ref{subsec:LF}. 
Larger samples over wider survey areas are required to better constrain the Pa$\beta$ AGN LF.

\subsection{Implication for Dusty Type 1 AGN} \label{subsec:dustAGN}

Our broad Paschen-line AGN sample contains dust-reddened and blue Type 1 AGNs without obvious selection biases. This advantage allows us to investigated the physical properties of dusty Type 1 AGNs. In this subsection, we explore whether the dust reddening is related to intrinsic AGN properties.

\begin{figure}[t]
\epsscale{1.2}
\plotone{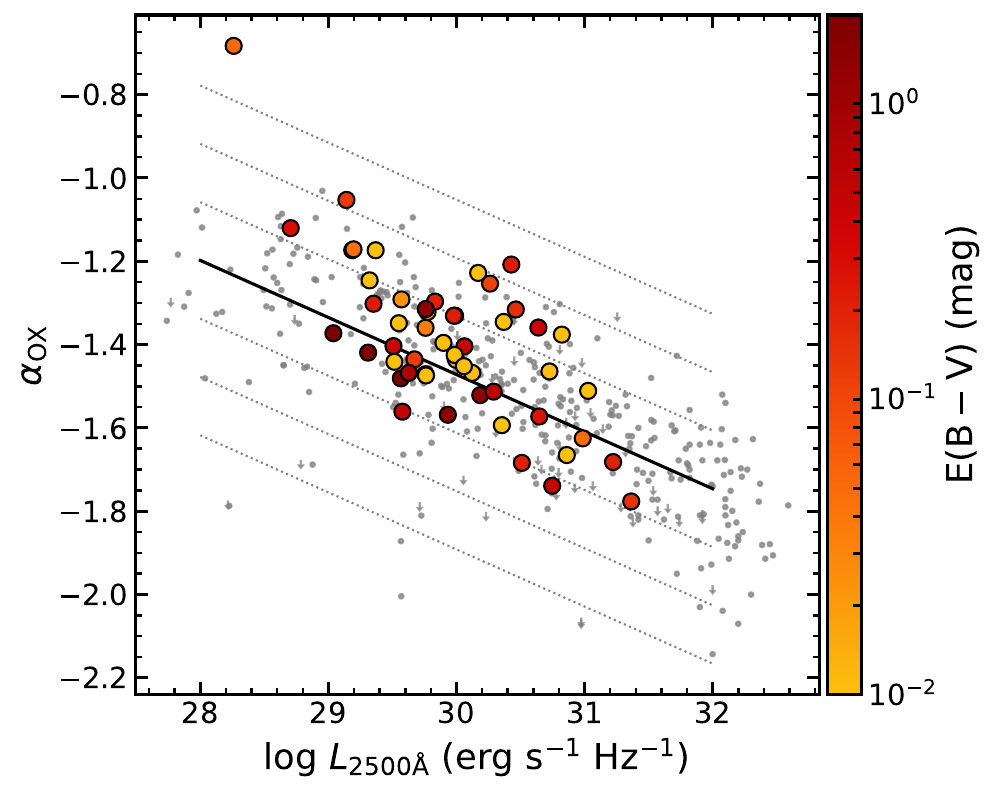}
\caption{Plot of $\alpha_{\mathrm{OX}}$ versus UV luminosity at rest-frame 2500 \AA. Our broad Paschen-line AGNs are shown as circles color-coded by $E(B-V)$ values. The typical errors for $\alpha_{\mathrm{OX}}$ and log $L_{2500\,\text{\AA}}$ are 0.04 and 0.11 dex, respectively. The $\alpha_{\mathrm{OX}}-L_{2500\,\text{\AA}}$ relation from \cite{steffen2006a} is shown as the solid line, with the dotted lines representing the 1, 2, 3 $\sigma$ scatter of the relation. The small grey dots and arrows represent the quasar sample of \cite{steffen2006a}. 
 \label{fig:alphaox}}
\end{figure}

The UV-to-X-ray spectral slope of AGNs is defined as
$\alpha_{\mathrm{OX}} = \log (L_{\mathrm{2\,keV}} / L_{2500\,\text{\AA}}) / \log (\nu_{\mathrm{2\,keV}} / \nu_{2500\,\text{\AA}})$ \citep{Tananbaum1979}, which serves as an indicator of the disk-corona ionizing state. Previous studies have established a tight correlation between $\alpha_{\mathrm{OX}}$ and the rest-frame monochromatic luminosity $L_{2500\,\text{\AA}}$ \citep[e.g.,][]{steffen2006a,lusso2010}. We compute $\alpha_{\mathrm{OX}}$ for broad Paschen-line AGNs with X-ray detections. 
The calculated $\alpha_{\mathrm{OX}}$ are presented in Appendix~\ref{sec:alllist}.
The rest-frame $L_{\mathrm{2\,keV}}$ is derived directly from the observed X-ray flux, and $L_{2500\,\text{\AA}}$ is obtained from the dust-corrected AGN SED model described in Section~\ref{sed}. Figure~\ref{fig:alphaox} shows the results, color-coded by $E(B-V)$ to indicate the dust-reddening level. 
Across the full reddening range, the majority of our AGNs consistently lie within the typical AGN region defined by the literature relation \citep{steffen2006a}. One outlier (ID638565) in the upper-left corner has an underestimated AGN $L_{2500\,\text{\AA}}$ caused by the poor image decomposition in a complex environment with nearby galaxies. Overall, the dust reddening is unrelated the ionizing conditions of Type 1 AGNs.

\begin{figure}[t]
\epsscale{1.2}
\plotone{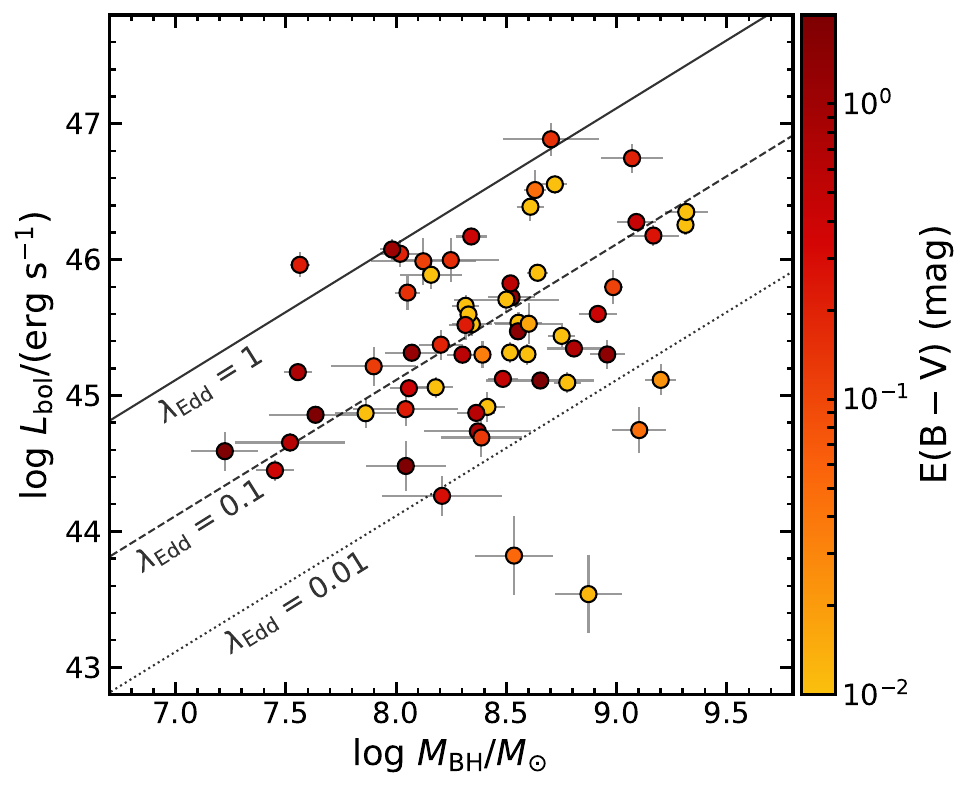}
\caption{Plot of bolometric luminosity versus $M_\mathrm{BH}$. Our broad Paschen-line AGNs are shown as circles color-coded by $E(B-V)$ values. The measurement errors are shown as the grey color bars. Grey solid, dashed and dotted lines represent the trends between $L_{\text{bol}}$ and $M_\mathrm{BH}$ with different Eddington ratio $\lambda_{\mathrm{Edd}}$ = 1, 0.1 and 0.01, respectively.
 \label{fig:edd}}
\end{figure}

To further probe the accretion properties, we analyze the Eddington ratio $\lambda_{\mathrm{Edd}}$ of our broad Paschen-line AGNs. We derive the bolometric luminosity $L_{\mathrm{bol}}$ by applying the bolometric correction from \cite{krawczyk2013}, using the same $L_{2500}$ as used for the $\alpha_{\mathrm{OX}}$ calculation. 
The calculated $L_{\mathrm{bol}}$ of the full sample are presented in Appendix~\ref{sec:alllist}.
The Eddington ratio is defined as $\lambda_{\mathrm{Edd}} = L_{\mathrm{bol}} / L_{\mathrm{Edd}}$, where $L_{\mathrm{Edd}} = 1.3 \times 10^{38} (M_{\mathrm{BH}}/M_{\odot})\ \mathrm{erg\ s^{-1}}$. Figure~\ref{fig:edd} shows $L_{\mathrm{bol}}$ versus $L_{\mathrm{Paschen}}$-based $M_\mathrm{BH}$, with $\lambda_{\mathrm{Edd}}$ indicated by the dashed lines. Two outliers (ID638565, ID502361) with log $L_{\mathrm{bol}}/(\mathrm{erg\ s^{-1}})$ smaller than 44 also suffer from poor image decomposition due to nearby galaxies. We find no significant trend against dust reddening, suggesting that dust reddening is not linked to the accretion state.
 
We find no intrinsic differences in the above properties across our broad Paschen-line AGN sample that spans a wide range of the dust-reddening level. This result suggests that dust reddening in Type 1 AGNs is more likely caused by the line-of-sight obscuration rather than different physical states of the central AGN engine. This scenario fits the dusty Type 1 AGN into the AGN unification framework. The dust reddening may arise from the clumpy dusty gas in the AGN torus or from dust on host-galaxy scales, which redden the AGN SED without fully obscuring the broad emission lines. Larger NIR-selected AGN samples in the future will provide more insights of this Type 1 AGN population.

\section{Summary} \label{sec:summary}

We have efficiently identified a sample of broad Pa$\alpha$ and Pa$\beta$ AGNs in the COSMOS-3D field using JWST WFSS data. The sample consists of 44 Pa$\alpha$ at $z = 1.1-1.7$ and 18 Pa$\beta$ at $z = 2.1-2.8$. To obtain robust, Paschen-based $M_{\mathrm{BH}}$ estimators, we have constructed a calibration sample with DESI optical spectra covering the \ion{Mg}{2} emission line. We performed a comprehensive analysis of the broad Paschen-line AGNs, including measurements of their spectral properties, host-galaxy contributions, and dust attenuation.

Based on the above data, we have derived new $M_{\mathrm{BH}}$ estimators for the Pa$\alpha$ and Pa$\beta$ AGNs at cosmic noon. We provide three types of $M_{\mathrm{BH}}$ estimators using Paschen-line luminosity and the AGN continuum luminosity at 1 and 2 $\mu$m. We removed the host-galaxy contamination from the continuum luminosities and highlighted the necessity of the dust-attenuation correction at 1 $\mu$m. Our $M_{\mathrm{BH}}$ estimators show a good self-consistency and also consistent with previous results.

We have further constructed Type 1 AGN LFs using our Pa$\alpha$ and Pa$\beta$ AGN samples. The resulting LFs are 3--5 times higher than those derived from optically selected samples, demonstrating the improved completeness of the NIR broad-line AGN selection. We found no intrinsic differences in the physical properties across the dust-reddening range of our sample. This supports a scenario in which dusty Type 1 AGNs are face-on AGNs within the AGN unification model; they are reddened by line-of-sight dusts in the AGN torus or the host galaxy. The growing JWST database and future large telescopes such as Roman Space Telescope will provide extensive Paschen-line detections. 

\begin{acknowledgments}
We acknowledge support from the National Science Foundation of China (12225301).
This work is based on observations made with the NASA/ESA Hubble Space Telescope and NASA/ESA/CSA James Webb Space Telescope. The data were obtained from the Mikulski Archive for Space Telescopes at the Space Telescope Science Institute, which is operated by the Association of Universities for Research in Astronomy, Inc., under NASA contract NAS 5-03127 for JWST. These observations are associated with programs 1635, 1727, 1810, 1837, 1933, 2321, 2514, 3990, 5398, 5893, 6368, 6434, and 6585. 
\end{acknowledgments}

\facilities{JWST, HST, Subaru (HSC), Mayall (DESI).}
\software{{\tt\string Astropy} \citep{astropy}, {\tt\string Numpy} \citep{numpy}, {\tt\string Scipy} \citep{scipy}, {\tt\string Matplotlib} \citep{matplotlib}, {\tt\string SExtractor} \citep{Bertin_1996}, {\tt\string PSFEx} \citep{bertin2013}, {\tt\string T-PHOT} \citep{Merlin2016}, {\tt\string SpectRes} \citep{carnall2017}, {\tt\string GalfitM} \citep{haussler2022a}, {\tt\string statmorph} \citep{statmorph}, {\tt\string photutils} \citep{larry_bradley_2024}, {\tt\string emcee} \citep{foreman-mackey2013}.}

\bibliography{ms}
\bibliographystyle{aasjournal}

\appendix

\section{Image Decomposition Results of Paschen-line AGN Sample}\label{sec:show_images}

\begin{figure*}
\epsscale{1}
\plotone{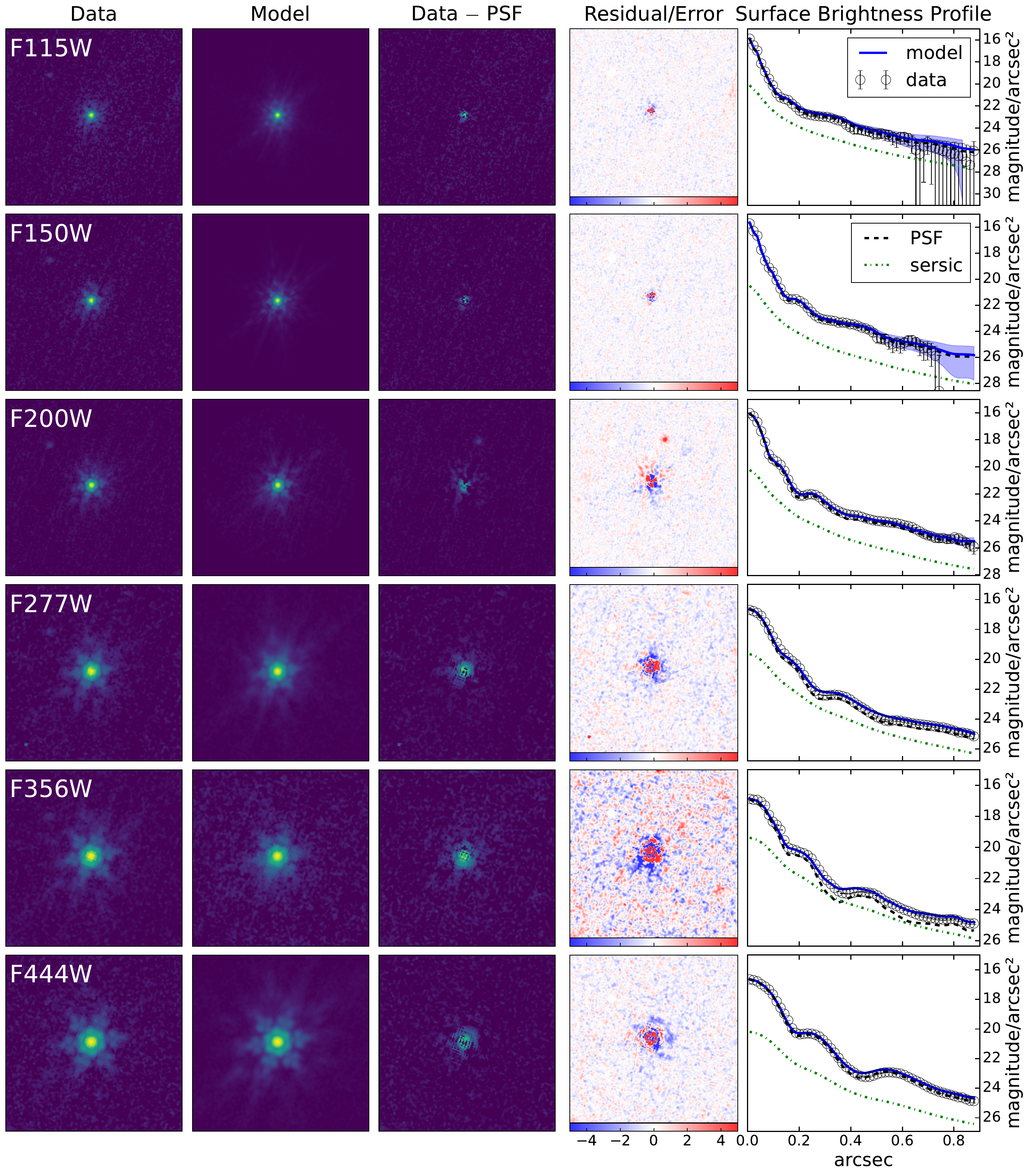}
\caption{An example (ID691466) of JWST image decomposition.
The size of each image is 150 pixels (4.5 arcsec) on a side. The five columns from left to right display the following information: (1) the observed JWST image, (2) the best-fitting model consisting of a PSF plus a S\'{e}rsic host galaxy, (3) the PSF-subtracted image, (4) the fitting residual image divided by the error image and (5) the surface brightness profiles. 
In column (5), we present the total model profiles (blue solid line), the PSF profiles (black dashed line), and the S\'{e}rsic profiles (green dotted line). The observed data points are shown with vertical error bars representing the 1$\sigma$ uncertainties. This image decomposition procedure works well for our broad Paschen-line AGNs.
 \label{fig:decomp1}}
\end{figure*}

\begin{figure*}
\epsscale{1}
\plotone{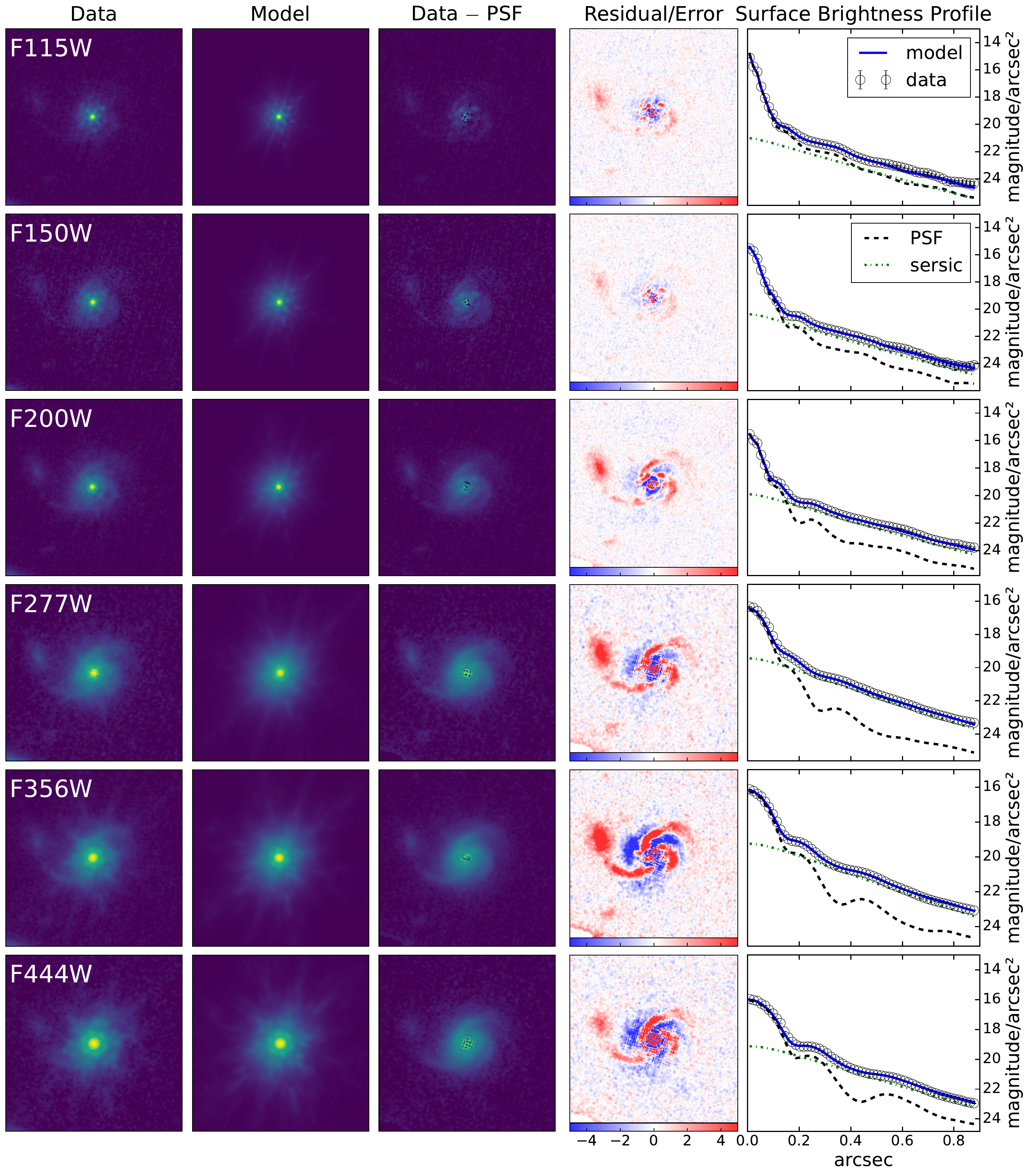}
\caption{An example (ID157997) of JWST image decomposition. Similar to Figure \ref{fig:decomp1}. Our image decomposition reliably separates the AGN and host-galaxy components for broad Paschen-line AGNs.
 \label{fig:decomp2}}
\end{figure*}

We present two examples of the image decomposition results for our broad Paschen-line AGNs. Figure~\ref{fig:decomp1} shows the decomposition of ID691466, which has AGN fractions of 0.75 and 0.85 at rest-frame 1 $\mathrm{\mu}$m and 2 $\mathrm{\mu}$m, respectively. The dominated PSF component is well modeled. Figure~\ref{fig:decomp2} shows the decomposition of ID157997, with AGN fractions of 0.49 and 0.53 at rest-frame 1 $\mathrm{\mu}$m and 2 $\mathrm{\mu}$m. After subtracting the PSF component, clear spiral-arm structures are revealed in the host galaxy.

\section{Measured Properties of the Paschen-line AGN sample}\label{sec:alllist}
We present the measured properties of our Paschen-line AGN sample in Table~\ref{tab:alllist}. The full table is available in machine-readable format online.

\begin{deluxetable*}{cccc}
\tablecaption{Measured Properties of the Paschen-line AGNs \label{tab:alllist}}
\tablewidth{20pt}
\tablehead{\colhead{Column} &\colhead{Format} &\colhead{Units} &\colhead{Description}}
\startdata
ID & int &  & Target ID in COSMOS2025 catalog \\
RA & float & deg & Right ascension (J2000) \\
DEC & float & deg & Declination (J2000) \\
Z & float &  & Spectroscopic redshift \\
LINE\_NAME & str &  & Paschen line name\\
F444W\_MAG & float & mag & Magnitude of JWST F444W band \\
F444W\_MAGERR & float & mag & Uncertainty of F444W\_MAG\\
FWHM & float & km s$^{-1}$ & Full width at half maximum of the Paschen line \\
FWHM\_ERR & float & km s$^{-1}$ & Uncertainty of FWHM \\
LOGL\_PA & float & erg s$^{-1}$ & Luminosity of the broad Paschen line \\
LOGL\_PA\_ERR & float & erg s$^{-1}$ & Uncertainty of LOGL\_PA \\
LOGL1um\_AGN & float & erg s$^{-1}$ & AGN luminosity at $1\,\mu$m \\
LOGL1um\_AGN\_ERR & float & erg s$^{-1}$ & Uncertainty of LOGL1um\_AGN \\
LOGL2um\_AGN & float & erg s$^{-1}$ & AGN luminosity at $2\,\mu$m \\
LOGL2um\_AGN\_ERR & float & erg s$^{-1}$ & Uncertainty of LOGL2um\_AGN \\
FRAC\_AGN\_1um & float &  & AGN fraction at $1\,\mu$m \\
FRAC\_AGN\_2um & float &  & AGN fraction at $2\,\mu$m \\
LOG\_Lbol\_AGN & float & erg s$^{-1}$ & AGN bolometric luminosity \\
LOG\_Lbol\_AGN\_ERR & float & erg s$^{-1}$ & Uncertainty of LOG\_Lbol\_AGN \\
EBV & float & mag & AGN dust reddening $E(B-V)$ \\
EBV\_ERR & float & mag & Uncertainty of EBV \\
alphaOX & float &  & UV-to-X-ray spectral slope $\alpha_{\mathrm{OX}}$ \\
alphaOX\_ERR & float &  & Uncertainty of alphaOX\\
\enddata
\tablecomments{This table is available in its entirety in machine-readable format.}
\end{deluxetable*}

\end{document}